\documentclass[twocolumn,superscriptaddress,showpacs]{revtex4}
\usepackage{graphicx}

\newcommand{\R}{\textbf{R}}

\begin{document}

\title{Nonlocal homogenization theory in metamaterials: effective electromagnetic spatial dispersion and artificial chirality}

\author{Alessandro Ciattoni}
\affiliation{Consiglio Nazionale delle Ricerche, CNR-SPIN, Via Vetoio 10, 67100 L'Aquila, Italy}
\author{Carlo Rizza}
\affiliation{Dipartimento di Scienza e Alta Tecnologia, Universit\`a dell'Insubria, Via Valleggio 11, 22100 Como, Italy}
\affiliation{Consiglio Nazionale delle Ricerche, CNR-SPIN, Via Vetoio 10, 67100 L'Aquila, Italy}

\begin{abstract}
We develop, from first principles, a general and compact formalism for predicting the electromagnetic response of a metamaterial with non-magnetic inclusions in the long wavelength limit, including spatial dispersion up to the second order. Specifically, by resorting to a suitable multiscale technique, we show that medium effective permittivity tensor and the first and second order tensors describing spatial dispersion can be evaluated by averaging suitable spatially rapidly-varying fields each satysifing electrostatic-like equations within the metamaterial unit cell. For metamaterials with negligible second-order spatial dispersion, we exploit the equivalence of first-order spatial dispersion and reciprocal bianisotropic electromagnetic response to deduce a simple expression for the metamaterial chirality tensor. Such an expression allows us to systematically analyze the effect of the composite spatial symmetry properties on electromagnetic chirality. We find that even if a metamaterial is geometrically achiral, i.e. it is indistinguishable from its mirror image, it shows pseudo-chiral-omega electromagnetic chirality if the rotation needed to restore the dielectric profile after the reflection is either a $0^\circ$ or $90^\circ$ rotation around an axis orthogonal to the reflection plane. These two symmetric situations encompass two-dimensional and one-dimensional metamaterials with chiral response. As an example admitting full analytical description, we discuss one-dimensional metamaterials whose single chirality parameter is shown to be directly related to the metamaterial dielectric profile by quadratures.
\end{abstract}


\maketitle

\section{Introduction}
Designing the electromagnetic response of an artificial medium is one of the main target of modern photonics and metamaterial science is probably the most important research field based on such skill. Basically the design is made possible by the physical fact that the electromagnetic field, when traveling within a nonhomogeneous medium with sub-wavelength features, is not able to follow its spatial rapidly-varying details so that the field only experiences the effect of an averaged or effective medium. A number of different homogenization approaches have been developed for predicting the effective medium electromagnetic response and they exploit different and suitable approximation schemes. The simplest homogenization technique deals with the retrieval of the effective parameters from the scattering properties of the medium \cite{OBrien,Smith1,Smith2,Chennn,Menzel,Plumm1,Simovs,Karama} and it is based on postulating the equivalence between a complex metamaterial array and a uniform slab of same thickness with unknown constitutive parameters. Another homogenization technique is the field averaging method which is based on the averaging of the electromagnetic field in a metamaterial unit cell \cite{Pendry,Smith3,Smith4,Tsuker,Ouchet} and, in analogy with the retrieval technique, it is a numerical method for the determination of effective parameters. In addition to numerical methods, mean-field homegenization theories are available where the effective parameters are evaluated from the distribution of the underlying metamaterial inclusions. Examples of such techniques are those exploiting Lorentz \cite{Ishima}, Clausius-Mossotti \cite{Belov1}, Maxwell-Garnett \cite{Lambbb} approximations, or based on multipolar expansion \cite{Aluuuu} and source-driven approach \cite{Silvei}. Spatial periodicity and rapidly-varying spatial scales are the two basic ingredients of each metamaterial homogenization approach. Two different homogenization techniques are obtained by assuming one of these two ingredients and subsequently incorporating the other. Therefore, starting from the photonic-crystal description of the structure where the spatial periodicity is fully taken into account, the effective medium response can be extracted in the long wavelength regime \cite{Taoooo,Dattaa,Halevi,Ortizz,Reyes1,PerezH,Reyes2}. Conversely, the spatial rapidly-varying metamaterial features allows an asymptotic multi-scale analysis of the sample electromagnetic response which combined with the array periodicity yields the effective medium response \cite{Felba0,Ouche1,Felbac,Chenn2,Rizza1,Rizza2,Rizza3}.

Even though the metamaterial inclusions pattering has a spatial scale much smaller than the radiation wavelength, such two scales are generally not so different to allow a description of the effective medium response only comprising effective dielectric permittivity and magnetic permeability tensors. For this reason, the effective medium generally shows an additional nonlocal respose \cite{Belov2,Gerken,Silve2,Elserr,Pollar,Chebyk,Orlovv} yielding spatial dispersion. It is well-known that the nonlocal first order contribution (i.e. containing first order spatial derivatives of the electric field) is equivalent to a reciprocal bianisotropic response \cite{Serdyu} whereas the second order contribution can be partially interpreted as a correction to the effective magnetic permeability tensor \cite{Serdyu,Menze1}, a phenomenon which is known as artificial or optical magnetism \cite{Pendry,Iwanag,Tserke,Chernn,Ginnnn}. If the effective nonlocality is weak, the effective medium response is adequately described by reciprocal bianisotropic constitutive relations where the chirality tensor accounts for the strength of magnetic and electric polarizations coupling. An efficient way for observing the effect of electromagnetic chirality is considering metamaterials whose underlying constituents' pattering exhibits chiral asymmetry, i.e. its mirror image cannot be superposed onto it, and theoretical and experimental investigations have been performed both in three-dimensional \cite{Jeli01,Gansel,Plum01,Wang01} and in two-dimensional configurations \cite{Papako,Fedoto,Singhh,Zhangg,Yeeeee}. Chiral metamaterials have attracted a good deal of attention since they can yield giant optical activity, asymmetric transmission \cite{Menze2,Liiiii}, repulsive Casimir force \cite{Zhaooo}, backward waves \cite{Trety3,Chern1} and negative refractive index \cite{Pendr2,Trety1,Trety2,Monzon}.

In this paper we theoretically investigate the electromagnetic response in the long wavelength regime of a periodic composite medium, whose inclusions are non-magnetic, and we carry out the analysis by including spatial dispersion up to the second order. Specifically, by using the ratio $\eta$ between the composite periodicity and the wavelength as an asymptotic expansion parameter, we exploit a general multiscale technique for separating the fast matter scale from the slow radiation one and to extract the average effect of the composite on the electromagnetic field. We fully develop the analysis up to the second order in $\eta$ and we find that the obtained effective medium response has a $\eta^0$ order reproducing the know results of standard homogenization approaches and $\eta^1$ and $\eta^2$ orders accounting for first and second order effective medium nonlocality, respectively. The effective dielectric permittivity tensor (of rank two) arising from the zeroth order is obtained by averaging over the metamataterial unit cell specific fast-varying fields satisfying electrostatic-like equations. The main result of the present paper is that this very simple procedure still works for evaluating the two tensors (of ranks three and four, respectively) describing effective medium nonlocality up to the second order and the corresponding electrostatic-like equations for the suitable fast-varying fields to be averaged are here derived and discussed. As a consequence we obtain a simple scheme for evaluating the effective medium response up to the second order which is based on solving a small number of Poisson equations on the metamaterial unit cell. In particular we obtain a very compact expression for the three-rank tensor describing first order spatial nonlocality which remarkably reveals that it can be evaluated by using the same fast-varying fields appearing in the zeroth order description, thus avoiding the requirement of solving additional electrostatic-like equations. Therefore the ensuing procedure for deducing the medium response up to the first order is remarkably simple and compact. For metamaterials with negligible second-order spatial dispersion, we combine the obtained description with the equivalence of first-order spatial dispersion and reciprocal bianisotropic electromagnetic response to obtain a compact and simple expression for the effective medium chirality tensor. Using such expression we investigate the impact of the composite patterning symmetry on electromagnetic chirality. Specifically we obtain that if the composite does not show chiral asymmetry, i.e. its mirror image can be superposed on it, the effective chiral tensor vanishes with two remarkable exceptions corresponding to the specific cases where the rotation needed to superpose the mirror image onto the structure is either a $0^\circ$ or $90^\circ$ rotation around an axis orthogonal to the reflection plane. In these two situations the composite is geometrically achiral and it nonetheless shows electromagnetic chirality whose chiral tensor turns out to be that of pseudo-chiral-omega medium. Such general result encompasses in particular the relevant cases of two-dimensional and recently considered one-dimensional metamaterials showing chiral response. As an example admitting full analytical description, we discuss one-dimensional metamaterials whose single chirality parameter is shown to be directly related to the metamaterial dielectric profile by quadratures.

The paper is organized as follows. In Sec. II we discuss our approach to the effective medium description  of a periodic composite in the long wavelength regime including spatial nonlocality. This is done examining the general multiscale technique in Sec. II.A and applying it to the zeroth, first and second orders in Sections II.B, II.C and II.D, respectively. In Sec. III we sum up the results obtained in Sec. II and we stress the simplicity and compactness of the proposed approach. In Sec. IV we consider media whose second order contribution to spatial nonlocal response can be neglected and we focus on electromagnetic chirality. Specifically in Sec. IV.A we adopt the bianisotropic description of the medium to deduce a simple expression for the chirality tensor, in Sec. IV.B we examine the relation between geometric and electromagnetic chirality of metamaterials and in Sec. IV.C we specialize such general analysis to the cases of two-dimensional and one-dimensional chiral metamaterials. In Sec. V we focus on one-dimensional chiral metamaterials and we obtain a closed-form expression for the single parameter ruling electromagnetic chirality of such media. In Sec. VI we draw our conclusions.

\section{Homogenization theory}
\subsection{Electromagnetic field multiscale analysis}
Let us consider propagation of a monochromatic electromagnetic field through an unbounded metamaterial whose underlying non-magnetic inclusions (both metal and dielectric) are arranged on a lattice whose period is much smaller than the radiation wavelength $\lambda$. The electric $\textbf{E}$ and magnetic $\textbf{H}$ field amplitudes satisfy Maxwell equations
\begin{eqnarray} \label{Maxwell}
\nabla \times \textbf{E} &=& i \omega \mu_0 \textbf{H}, \nonumber \\
\nabla \times \textbf{H} &=& -i \omega \varepsilon_0 \varepsilon_r \textbf{E}
\end{eqnarray}
where time dependence $e^{-i\omega t}$ has been assumed and the medium relative dielectric constant $\varepsilon_r (\textbf{r})$  is a periodic complex function having the same periodicity of the metamaterial. Due to the rapidly varying dielectric spatial modulation it is convenient to introduce the parameter $\eta = d/\lambda$ where $d$ is the largest of the lattice basis vector lengths. Following the standard multiscale technique, it is possible to exploit the condition $\eta \ll 1$ to develop an asymptotic analysis of electromagnetic propagation. Accordingly, we introduce the fast spatial coordinates $\R = \textbf{r} / \eta$, regarded as independent from the slow coordinates $\textbf{r}$. Any field $f(\textbf{r},\R)$ can be decomposed as $f(\textbf{r},\R) = \overline{f}(\textbf{r})  + \tilde{f}(\textbf{r},\R)$ where the overline denote the spatial average over the metamaterial unit cell, i.e
\begin{equation}
\overline{f}(\textbf{r}) = \frac{1}{V} \int_C d^3 \R f(\textbf{r},\R)
\end{equation}
where $C$ is the unit cell and $V$ is its volume scaled by $\eta^3$, and the tilde denote the rapidly varying zero mean residual, i.e. $\tilde{f} = f - \overline{f} $. In our scheme the relative dielectric permittivity depends on the fast coordinates only so that we set
\begin{equation} \label{dielectric}
\varepsilon(\R) = \varepsilon_r(\eta \R)
\end{equation}
and it can be decomposed as $\varepsilon(\R) = \overline{\varepsilon} + \tilde{\varepsilon}(\R)$. Following the general multiscale technique \cite{Sander}, we set for the electromagnetic field amplitudes $\textbf{A}=\textbf{E},\textbf{H}$
\begin{equation} \label{two-scale}
\textbf{A} = \sum_{n=0}^{\infty} \left[\overline{\textbf{A}}_n (\textbf{r}) + \tilde{\textbf{A}}_n (\textbf{r},\R) \right] \eta^n,
\end{equation}
or, in other words, the field amplitudes are expanded in powers of $\eta$ and, for each order, the average on the unit cell (terms with an overbar and depending on the slow coordinates only) is separated from the zero mean residual (terms with an overtilde and depending on both slow and fast coordinates). After substituting Eqs.(\ref{two-scale}) into Eqs.(\ref{Maxwell}) and noting that $\nabla \rightarrow \nabla + \frac{1}{\eta} \nabla_\R$ it is possible to extract equations for each order and, for each of them, to separately balance the averaged contributions and zero mean residuals. As a result, for the averaged equations we obtain
\begin{eqnarray} \label{n-slowly}
\nabla \times \overline{\textbf{E}}_n &=& i \omega \mu_0 \overline{\textbf{H}}_n, \nonumber \\
\nabla \times \overline{\textbf{H}}_n &=& -i \omega \varepsilon_0 \left( \overline{\varepsilon} \overline{\textbf{E}}_n + \overline{\varepsilon \tilde{\textbf{E}}_n} \right)
\end{eqnarray}
whereas for the zero mean residual equations we have
\begin{eqnarray} \label{n-rapid}
\nabla_{\R} \times \tilde{\textbf{E}}_0 &=& \textbf{0} , \nonumber \\
\nabla_{\R} \times \tilde{\textbf{H}}_0 &=& \textbf{0} , \nonumber \\
\nabla_{\R} \times \tilde{\textbf{E}}_{n+1} &=& -\nabla \times \tilde{\textbf{E}}_n + i \omega \mu_0 \tilde{\textbf{H}}_n , \nonumber \\
\nabla_{\R} \times \tilde{\textbf{H}}_{n+1} &=& -\nabla \times \tilde{\textbf{H}}_n \nonumber \\
&& - i \omega \varepsilon_0 \left[ (\varepsilon-\overline{\varepsilon}) \overline{\textbf{E}}_n + \varepsilon \tilde{\textbf{E}}_n - \overline{ \varepsilon \tilde{\textbf{E}}_n} \right]
\end{eqnarray}
where $n=0,1,2,\dots$. Equations (\ref{n-slowly}) do not contain the fast coordinates $\R$ and are coupled to the rapidly varying field orders only through the averaged term $\overline{ \varepsilon \tilde{\textbf{E}}_n}$. Multiplying each of Eqs.(\ref{n-slowly}) for $\eta^n$ and summing over $n$ we obtain
\begin{eqnarray} \label{macroscopic}
\nabla \times \overline{\textbf{E}} &=& i \omega \mu_0 \overline{\textbf{H}}, \nonumber \\
\nabla \times \overline{\textbf{H}} &=& -i \omega \overline{\textbf{D}}
\end{eqnarray}
where $\overline{\textbf{E}} = \sum_{n=0}^\infty \overline{\textbf{E}}_n \eta^n$ and $\overline{\textbf{H}} = \sum_{n=0}^\infty \overline{\textbf{H}}_n \eta^n$ is the overall averaged electromagnetic field and
\begin{equation} \label{eff-disp}
\overline{\textbf{D}} = \varepsilon_0 \overline{\varepsilon} \overline{\textbf{E}} + \sum_{n=0}^\infty \overline{\Delta \textbf{D}}_n
\end{equation}
where
\begin{equation} \label{Dn}
\overline{\Delta \textbf{D}}_n = \varepsilon_0  \overline{ \varepsilon \tilde{\textbf{E}}_n}  \eta^n.
\end{equation}
Therefore the slowly varying electric and magnetic field amplitudes satisfy the macroscopic Maxwell equations (\ref{macroscopic}) with the slowly varying displacement vector $\overline{\textbf{D}}$ of Eq.(\ref{eff-disp}) which has two contributions, the former due to the spatial average of the dielectric profile and latter due to the dielectric modulation. The latter contribution is obtained by summing the spatial average of the rapidly varying fields $\tilde{\textbf{E}}_n$ multiplied by the dielectric permittivity arising from each order. Therefore, in order to obtain an effective medium description of the metamaterial response, such rapidly varying fields have to be related to the slowly varying ones through Eqs.(\ref{n-rapid}).

From the structure of Eqs.(\ref{n-rapid}) we note that the fields $\tilde{\textbf{E}}_{n+1}$ and $\tilde{\textbf{H}}_{n+1}$ have to be evaluated recursively from the knowledge of $\tilde{\textbf{E}}_{n}$ and $\tilde{\textbf{H}}_{n}$. In order to accomplish this task it is convenient to deduce equations involving the fields divergence. After applying the operator $\nabla_\R \cdot$ to both the third and fourth of Eqs.(\ref{n-rapid}) we obtain
\begin{eqnarray} \label{div}
\nabla \cdot \left( \nabla_\R \times \tilde{\textbf{E}}_n \right) &=& -i \omega \mu_0 \nabla_\R \cdot \tilde{\textbf{H}}_n, \nonumber \\
\nabla \cdot \left( \nabla_\R \times \tilde{\textbf{H}}_n \right) &=&  i \omega \varepsilon_0 \nabla_\R \cdot \left[ (\varepsilon - \overline{\varepsilon}) \overline{\textbf{E}}_n + \varepsilon \tilde{\textbf{E}}_n \right]
\end{eqnarray}
where we have used the identity $\nabla_\R \cdot ( \nabla \times \textbf{A} )= - \nabla \cdot (\nabla_\R \times \textbf{A})$. Setting $n=0$ and using the first and the second of Eqs.(\ref{n-rapid}), Eqs.(\ref{div}) become
\begin{eqnarray} \label{div0}
\nabla_\R \cdot \tilde{\textbf{H}}_0 &=& 0, \nonumber \\
\nabla_\R \cdot \left( \varepsilon \tilde{\textbf{E}}_0 \right) &=& - \left( \nabla_\R \varepsilon \right) \cdot \overline{\textbf{E}}_0.
\end{eqnarray}

Relabelling $n \rightarrow n+1$ in Eqs.(\ref{div}) and substituting the expressions for $\nabla_\R \times \tilde{\textbf{E}}_{n+1}$ and $\nabla_\R \times \tilde{\textbf{H}}_{n+1}$ from the third and fourth of Eqs.(\ref{n-rapid}) we obtain
\begin{eqnarray} \label{divn}
\nabla_\R \cdot \tilde{\textbf{H}}_{n+1} &=& -\nabla_\R \cdot \tilde{\textbf{H}}_n,
\nonumber \\
\nabla_\R \cdot \left( \varepsilon \tilde{\textbf{E}}_{n+1} \right) &=&
- \nabla \cdot \left[ (\varepsilon-\overline{\varepsilon})  \overline{\textbf{E}}_n + \varepsilon \tilde{\textbf{E}}_n - \overline{\varepsilon \tilde{\textbf{E}}_n} \right]
\nonumber \\
&& - \left(\nabla_\R \varepsilon \right) \cdot \overline{\textbf{E}}_{n+1}
\end{eqnarray}
for $n=0,1,\dots$. Equations (\ref{n-rapid}) together with Eqs.(\ref{div0}) and (\ref{divn}) can be used to evaluate the rapidly varying fields of order $n+1$ once those of order $n$ are known and these fields are linearly dependent on the slowly-varying fields. Therefore, the discussed homogenization technique allows to obtain an effective medium description of the metamaterial respose up to the desired order in $\eta$. In this paper we will provide an effective medium description up to the second order.

\subsection{Zeroth order}
From the second of Eqs.(\ref{n-rapid}) and the first of Eqs.(\ref{div0}) we note that the field $\tilde{\textbf{H}}_0$ has no sources and therefore it vanishes. On the other hand the first of Eqs.(\ref{n-rapid}) and the second of Eqs.(\ref{div0}) state that the field $\tilde{\textbf{E}}_{0}$ is conservative and it is produced by the spatial modulation of $\varepsilon$ and the slowly varying field $\overline{\textbf{E}}_{0}$. Therefore we have
\begin{eqnarray} \label{0-order}
\tilde{\textbf{H}}_{0} &=& \textbf{0}, \nonumber \\
\tilde{\textbf{E}}_{0} &=& \hat{\textbf{e}}_i \left(\partial_i f_j\right) \overline{E}_{0j}
\end{eqnarray}
where the sum is hereafter understood over repeated indices, $\hat{\textbf{e}}_i$ is the unit vector along the $i$-th direction, $\partial_i$ is the partial derivative along $X_i = \hat{\textbf{e}}_i \cdot \R$, $\overline{E}_{0j}= \hat{\textbf{e}}_j \cdot \overline{\textbf{E}}_{0}$ and the functions $f_j$ satisfy the equations
\begin{equation} \label{f0}
\nabla_\R \cdot \left( \varepsilon \nabla_\R f_j \right) = -\partial_j \varepsilon.
\end{equation}
where $j=1,2,3$. Note that the source term of Eq.(\ref{f0}) has the same metamaterial periodicity and its spatial average vanishes, i.e. $\overline{\partial_j \varepsilon} = 0$, so that the fields $f_j$ are periodic functions with the same metamaterial periodicity and they are defined up to an arbitrary constant (see Appendix A). Inserting the second of Eqs.(\ref{0-order}) into Eq.(\ref{Dn}) with $n=0$ we obtain
\begin{equation} \label{D0}
\overline{\Delta \textbf{D}}_0 = \varepsilon_0 \hat{\textbf{e}}_i \Delta \varepsilon^{(eff)}_{ij} \overline{E}_{0j}.
\end{equation}
where $\Delta \varepsilon^{(eff)}_{ij} = \overline{ \varepsilon \partial_i f_j}$ so that the zeroth order contribution to the displacement vector of the effective medium amounts to a tensor correction to the average dielectric permittity. This result agrees with Ref.\cite{Felbac} where the authors consider the homogenization of a two-dimensional structure up to the zeroth order in $\eta$. Note that, as shown in appendix B, the rank-two tensor $\Delta \varepsilon^{(eff)}_{ij}$ satisfies the relation $\Delta \varepsilon^{(eff)}_{ij} = \Delta \varepsilon^{(eff)}_{ji}$ and this is the correct symmetry property of the correction to the dielectric permittivity tensor, in agreement with the Onsager symmetry principle \cite{Landau}. After symmetrizing the pair of indices $ij$, i.e. by setting $\Delta \varepsilon^{(eff)}_{ij} = \frac{1}{2} \left( \Delta \varepsilon^{(eff)}_{ij} + \Delta \varepsilon^{(eff)}_{ji} \right)$, we get
\begin{equation} \label{depeff}
\Delta \varepsilon^{(eff)}_{ij} = -\delta_{ij} \overline{ \varepsilon} + \frac{1}{2} \overline{ \left(Q_{ij}+Q_{ji}\right) }.
\end{equation}
where we have set
\begin{equation} \label{Q}
Q_{ij}= \varepsilon \left(\delta_{ij} + \partial_i f_j \right),
\end{equation}
for later convenience. It is worth noting that the expression for $Q_{ij}$ only involves spatial derivatives of the fields $f_j$ so that, at the zeroth order, the undefined additive constant in $f_j$ (see Appendix A) do not affect the effective correction to the permittivity tensor of Eq.(\ref{depeff}). In addition, as a consequence of Eq.(\ref{f0}) we have the divergenceless condition
\begin{equation} \label{div-less}
\partial_i Q_{ij} = \partial_j \varepsilon + \nabla_\R \cdot \left( \varepsilon \nabla_\R f_j \right) = 0.
\end{equation}

\subsection{First order}
In order to obtain the equations for the fast varying fields of the first order we use the third and fourth of Eqs.(\ref{n-rapid}) and Eqs.(\ref{divn}) with $n=0$ thus getting
\begin{eqnarray} \label{1-order-eqs}
\nabla_\R \cdot \tilde{\textbf{H}}_1 &=& 0,
\nonumber \\
\nabla_{\R} \times \tilde{\textbf{H}}_1 &=&  - i \omega \varepsilon_0 \left[ \hat{\textbf{e}}_i \left(Q_{ij}-\overline{Q_{ij}}\right) \overline{E}_{0j}\right], \nonumber \\
\nabla_\R \cdot \left( \varepsilon \tilde{\textbf{E}}_1 \right) &=& - \left(Q_{ij}-\overline{Q_{ij}}\right) \frac{\partial \overline{E}_{0j}}{\partial x_i}
-(\partial_i \varepsilon ) \overline{E}_{1i},
\nonumber \\
\nabla_{\R} \times \left( \tilde{\textbf{E}}_1 - \hat{\textbf{e}}_i f_j \frac{\partial \overline{E}_{0j}}{\partial x_i}  \right) &=& \textbf{0}
\end{eqnarray}
where we have used Eqs.(\ref{0-order}) and (\ref{Q}) and we have exploited the evident relation $\nabla \times \tilde{\textbf{E}}_0 = - \nabla_{\R} \times \left( \hat{\textbf{e}}_i f_j \frac{\partial \overline{E}_{0j}}{\partial x_i}  \right)$. Note that we use the symbol $\partial/\partial x_i$ to label the partial derivative operators with respect the slowly varying coordinates $x_i$ in order to avoid confusion with $\partial_i = \partial / \partial X_i$ which we use for partial derivative operators with respect the fast varying coordinates.

The first and the second of Eqs.(\ref{1-order-eqs}) provide the field $\tilde{\textbf{H}}_1$ since
\begin{equation}
\nabla_\R \cdot \left( \hat{\textbf{e}}_i Q_{ij} \overline{E}_{0j}\right) = \left( \partial_i Q_{ij} \right) \overline{E}_{0j} = 0,
\end{equation}
where Eq.(\ref{div-less}) has been used, is the compatibility condition of the system. Therefore we set
\begin{equation} \label{H1}
\tilde{\textbf{H}}_1 = \frac{1}{i \omega \mu_0} \left[ \nabla_{\R} \times \left( \hat{\bf e}_i A_{ij}\right)  \right]  \overline{E}_{0j}
\end{equation}
where the functions $A_{ij}$ satisfy the magnetostatic equations
\begin{eqnarray} \label{magneto}
\partial_i A_{ij} &=& 0, \nonumber \\
\nabla_{\R}^2 A_{ij} &=& - k_0^2 \left( Q_{ij} - \overline{Q_{ij}} \right)
\end{eqnarray}
where $k_0 = \omega/c$, which have to be solved with the prescritions that $A_{ij}$ has the metamaterial periodicity and it is defined up to an arbitrary constant (see Appendix A). Note that if $A_{ij}$ satisfies the second of Eqs.(\ref{magneto}) then
\begin{equation}
\nabla_{\R}^2 \left( \partial_i A_{ij} \right) =  - k_0^2  \partial_i Q_{ij} = 0
\end{equation}
where Eq.(\ref{div-less}) has been used and this implies that $\partial_i A_{ij} = const.$ since the constant is the only harmonic function which has the metamaterial periodicity. In addition $\partial_i A_{ij}$ has vanishing average since it is the derivative of a periodic function and therefore $\partial_i A_{ij} = 0$. Therefore the first of Eqs.(\ref{magneto}) is automatically satisfied and has not to be additionally required.

The third and the fourth of Eqs.(\ref{1-order-eqs}) provide the field $\tilde{\textbf{E}}_1$ and it is easily seen that it can be expressed as
\begin{equation} \label{E1}
\tilde{\textbf{E}}_1 = \hat{\textbf{e}}_i \left[ \left( \partial_i f_j \right) \overline{E}_{1j} + \left(  \delta_{ir} f_j+ \partial_i W_{rj} \right) \frac{\partial \overline{E}_{0j}}{\partial x_r}  \right]
\end{equation}
where the functions $W_{rj}$ satisfy the equation
\begin{equation} \label{f1}
\nabla_\R \cdot \left( \varepsilon \nabla_\R W_{rj} \right) = -\partial_r \left( \varepsilon f_j\right) - \left( Q_{rj} - \overline{Q_{rj}} \right).
\end{equation}
It is important noting that field $\tilde{\textbf{E}}_1$ has, by its very definition, zero average and, from Eq.(\ref{E1}) this forces the undefined additive constant of $f_j$ to be set equal to zero, i.e. in such a way that
\begin{equation}
\overline{f_j} = 0.
\end{equation}
Note that the source term of Eq.(\ref{f1}), in its right hand side, has vanishing spatial average so that the equation is structurally equivalent to Eq.(\ref{f0}) and therefore $W_{rj}$ has the metamaterial periodicity and it is defined up to an additive constant (see Appendix A). Inserting Eq.(\ref{E1}) into Eq.(\ref{Dn}) with $n=1$ we obtain
\begin{equation} \label{D1}
\overline{\Delta \textbf{D}}_1 = \varepsilon_0 \hat{\textbf{e}}_i \left[ \eta \Delta \varepsilon^{(eff)}_{ij} \overline{E}_{1j} +
 \alpha^{(eff)}_{ijr} \frac{\partial \overline{E}_{0j}}{\partial x_r} \right]
\end{equation}
where $\alpha^{(eff)}_{ijr} = \eta \overline{ \varepsilon \left(  \delta_{ir} f_j+ \partial_i W_{rj} \right)}$ so that the first order contribution to the displacement vector of the effective medium provides a tensor correction to the average dielectric permittivity which has the same structure as its zeroth order counterpart (see Eq.(\ref{D0})) and a contribution which is linear in the derivatives of the slowly varying zeroth order electric field. This former contribution is responsible for the first order nonlocal part of the effective medium dielectric response.

The rank-three tensor $\alpha^{(eff)}_{ijr}$ can be conveniently rewritten as (see Appendix B)
\begin{equation} \label{alpha}
\alpha^{(eff)}_{ijr} = \eta \overline{ \left( Q_{ri} f_j - Q_{rj} f_i \right) }
\end{equation}
which reveals that the nonlocal contribution of the effective medium response up to the first order can be predicted directly from the knowledge of the functions $f_i$ thus avoiding to solve Eqs.(\ref{f1}). Note that Eq.(\ref{alpha}) also shows that
\begin{equation} \label{antsym}
\alpha^{(eff)}_{ijr} = - \alpha^{(eff)}_{jir}
\end{equation}
which is the correct antisymmetric property, in agreement with the Onsager symmetry principle, for the third order tensor associated with first order spatial nonlocality of the effective medium response \cite{Landau}.

\subsection{Second order}

The equations for the fast varying electric field of the second order are obtained from the third of Eqs.(\ref{n-rapid}) and the second of Eqs.(\ref{divn}) with $n=1$ so that
\begin{eqnarray} \label{2-order-eqs}
\nabla_\R \cdot \left( \varepsilon \tilde{\textbf{E}}_2 \right) &=&
- \left(Q_{ij} - \overline{Q_{ij}} \right) \frac{\partial \overline{E}_{1j}}{\partial x_i}
- \left(P_{irj} - \overline{P_{irj}} \right) \frac{\partial^2 \overline{E}_{0j}}{\partial x_i \partial x_r} \nonumber \\ &&
- (\partial_i \varepsilon) \overline{E}_{2i}, \nonumber \\
 \nabla_{\R} \times \tilde{\textbf{E}}_2 &=& \nabla_{\R} \times \left[ \hat{\textbf{e}}_i \left(
 f_j \frac{\partial \overline{E}_{1j}}{\partial x_i}  + W_{rj} \frac{\partial^2 \overline{E}_{0j}}{\partial x_i \partial x_r} + A_{ij} \overline{E}_{0j} \right) \right]  \nonumber \\
\end{eqnarray}
where we have used Eqs.(\ref{H1}) and (\ref{E1}), we have set
\begin{equation} \label{P}
P_{irj} = \varepsilon \left(  \delta_{ir} f_j + \partial_i W_{rj} \right)
\end{equation}
and we have used the evident relation $\nabla \times \tilde{\textbf{E}}_1 = - \nabla_{\textbf{R}} \times \left[ \hat{\textbf{e}}_i \left( f_j \frac{\partial \overline{E}_{1j}}{\partial x_i}  + W_{rj} \frac{\partial^2 \overline{E}_{0j}}{\partial x_i \partial x_r} \right) \right]$. Note that, since we are investigating the effective medium response up to the second order $\eta^2$, we are not considering the equations for the second order magnetic field $\tilde{\textbf{H}}_2$ since it provides contributions to the effective medium response only of order $\eta^3$.

It is easily seen that the field $\tilde{\textbf{E}}_2$ satisfying Eqs.(\ref{2-order-eqs}) is
\begin{eqnarray} \label{E2}
\tilde{\textbf{E}}_2 &=& \hat{\textbf{e}}_i  \left[ \left( \partial_i f_j \right) \overline{E}_{2j} + \left(  \delta_{ir} f_j+ \partial_i W_{rj} \right) \frac{\partial \overline{E}_{1j}}{\partial x_r}  \right. \nonumber \\
&+& \left.  \left(A_{ij}+ \partial_i v_j\right) \overline{E}_{0j} +  \left(  \delta_{is} W_{rj}+ \partial_i R_{srj} \right) \frac{\partial^2 \overline{E}_{0j}}{\partial x_s \partial x_r} \right] \nonumber \\
\end{eqnarray}
where the functions $R_{srj}$ and $v_j$ satisfy the equations
\begin{eqnarray} \label{R}
\nabla_\R \cdot \left( \varepsilon \nabla_\R R_{srj} \right) &=& -\partial_s \left( \varepsilon W_{rj} \right) - \left( P_{srj} - \overline{P_{srj}} \right), \nonumber \\
\nabla_\R \cdot \left( \varepsilon \nabla_\R v_j \right) &=& -\partial_i \left(\varepsilon A_{ij}\right).
\end{eqnarray}
It is important noting that field $\tilde{\textbf{E}}_2$ has, by its very definition, zero average and, from Eq.(\ref{E2}) this forces the undefined additive constants of $A_{ij}$ and $W_{rj}$ to be set equal to zero, i.e. in such a way that
\begin{eqnarray}
\overline{A_{ij}} &=& 0, \nonumber \\
\overline{W_{rj}} &=& 0.
\end{eqnarray}
Note that the source terms of Eqs.(\ref{R}) have vanishing spatial average so that such equations are structurally equivalent to Eq.(\ref{f0}) and therefore $R_{srj}$ and $v_j$  have the metamaterial periodicity and they are defined up to an arbitrary constant (see Appendix A). Inserting Eq.(\ref{E2}) into Eq.(\ref{Dn}) with $n=2$ we obtain
\begin{eqnarray} \label{D2}
\overline{\Delta \textbf{D}}_2 =
\varepsilon_0 \hat{\textbf{e}}_i &&
\left[ \eta^2 \Delta \varepsilon^{(eff)}_{ij} \overline{E}_{2j} +
\eta \alpha^{(eff)}_{ijr} \frac{\partial \overline{E}_{1j}}{\partial x_r} \right. \nonumber \\ &&+ \left.
\eta^2 \gamma^{(eff)}_{ij} \overline{E}_{0j} +
 \beta^{(eff)}_{ijsr} \frac{\partial^2 \overline{E}_{0j}}{\partial x_s \partial x_r} \right]
\end{eqnarray}
where $\gamma^{(eff)}_{ij} = \overline{\varepsilon \left(A_{ij}+ \partial_i v_j\right)}$ and $\beta^{(eff)}_{ijsr} = \frac{\eta^2}{2} \overline{\varepsilon \left(  \delta_{is} W_{rj} + \delta_{ir} W_{sj} + \partial_i R_{srj} + \partial_i R_{rsj} \right)}$. Note that in the fourth term in Eq.(\ref{D2}), $\beta^{(eff)}_{ijsr}$ has been symmetrized with respect the indices $sr$ since it is contracted with the symmetric mixed partial derivatives tensor. From Eq.(\ref{D2}) we note that second order contribution to the displacement vector of the effective medium contains a correction to the effective dielectric permittivity tensor and a correction to the first order nonlocal response which which have the same structures as their zeroth and first order counterpart (see Eqs.(\ref{D0}) and (\ref{D1})). In addition it provides a novel contribution to the effective dielectric permittivity which is physically due to the effect of the rapidly varying first order magnetic field and a contribution which is linear in the mixed second order derivatives of the slowly varying zeroth order electric field, the latter being responsible for the second order nonlocal part of the effective medium dielectric response.

The rank-two tensor $\gamma^{(eff)}_{ij}$ and the rank-four tensor $\beta^{(eff)}_{ijsr}$ can be rewritten as (see Appendix B)
\begin{eqnarray} \label{gam-bet}
\gamma^{(eff)}_{ij} &=& \frac{1}{k_0^2} \overline{ ( \partial_s A_{ri})( \partial_s A_{rj}) }, \nonumber \\
\beta^{(eff)}_{ijsr} &=&
\frac{\eta^2}{4}\overline{ \left( Q_{ri} W_{sj} + Q_{si} W_{rj} + Q_{rj} W_{si} + Q_{sj} W_{ri} \right) } \nonumber \\ && -
\frac{\eta^2}{4}\overline{ \left[ f_i \left(P_{rsj} + P_{srj}\right) + f_j \left( P_{rsi} + P_{sri} \right) \right]}. \nonumber \\
\end{eqnarray}
so that the second order contribution to the effective medium response can be predicted without solving Eqs.(\ref{R}). It is worth noting that the relations $\gamma^{(eff)}_{ij} = \gamma^{(eff)}_{ji}$ and $\beta^{(eff)}_{ijrs} = \beta^{(eff)}_{jirs}$ hold and they are the correct symmetric properties, in agreement with the Onsager symmetry principle, for contributions to the effective dielectric tensor and for the rank-four tensor associated with second order spatial nonlocality \cite{Landau}, respectively.

\section{Effective medium response up to the second order}
The slowly varying electric and displacement fields up the second order are $\overline{\textbf{E}} = \overline{E}_i \textbf{e}_i =  \overline{\textbf{E}}_0  + \overline{\textbf{E}}_1 \eta + \overline{\textbf{E}}_2 \eta^2$ and $\overline{\textbf{D}} = \overline{D}_i \textbf{e}_i =  \overline{\textbf{D}}_0  + \overline{\textbf{D}}_1 \eta + \overline{\textbf{D}}_2 \eta^2$ so that, inserting Eqs.(\ref{D0}), (\ref{D1}) and (\ref{D2}) into Eq.(\ref{eff-disp}) (whose series is truncated up to the second order)
and adding suitable higher order terms for restoring the field $\overline{E}_i$ in the nonlocal contributions, we obtain
\begin{eqnarray}  \label{eff-med}
\overline{D}_i &=&
\varepsilon_0 \left(
\varepsilon^{(eff)}_{ij} \overline{E}_j +
\alpha^{(eff)}_{ijr} \frac{\partial \overline{E}_j}{\partial x_r} +
\beta^{(eff)}_{ijrs} \frac{\partial^2 \overline{E}_j}{\partial x_r \partial x_s}
\right) \nonumber \\
\end{eqnarray}
where, from Eqs.(\ref{depeff}), (\ref{alpha}) and (\ref{gam-bet}), we have
\begin{eqnarray}  \label{eff-tens}
\varepsilon^{(eff)}_{ij} &=& \frac{1}{2} \overline{  \left( Q_{ij} + Q_{ji}  \right)} +  \frac{\eta^2}{k_0^2} \overline{( \partial_s A_{ri})( \partial_s A_{rj}) }, \nonumber \\
\alpha^{(eff)}_{ijr} &=&  \eta \overline{ \left( Q_{ri} f_j - Q_{rj} f_i \right) }, \nonumber \\
\beta^{(eff)}_{ijsr} &=&
\frac{\eta^2}{4}\overline{ \left( Q_{ri} W_{sj} + Q_{si} W_{rj} + Q_{rj} W_{si} + Q_{sj} W_{ri} \right) } \nonumber \\ && -
\frac{\eta^2}{4}\overline{ \left[ f_i \left(P_{rsj} + P_{srj}\right) + f_j \left( P_{rsi} + P_{sri} \right) \right]}. \nonumber \\
\end{eqnarray}
where
\begin{eqnarray} \label{eff-QP}
Q_{ij} &=& \varepsilon \left(\delta_{ij} + \partial_i f_j \right), \nonumber \\
P_{irj} &=& \varepsilon \left(  \delta_{ir} f_j + \partial_i W_{rj} \right).
\end{eqnarray}
In order to sum up the results obtained so far for clarity purposes, we here report Eq.(\ref{f0}), the second of Eq.(\ref{magneto}) and Eq.(\ref{f1}) for the fields $f_j$, $A_{ij}$ and $W_{rj}$ necessary for evaluating the effective medium tensors, namely
\begin{eqnarray}  \label{eff-eqs}
\nabla_\R \cdot \left( \varepsilon \nabla_\R f_j \right) &=& -\partial_j \varepsilon,  \nonumber \\
\nabla_{\R}^2 A_{ij} &=& - k_0^2 \left( Q_{ij} - \overline{Q_{ij}} \right),  \nonumber \\
\nabla_\R \cdot \left( \varepsilon \nabla_\R W_{rj} \right) &=& -\partial_r \left( \varepsilon f_j\right) - \left( Q_{rj} - \overline{Q_{rj}} \right),
\end{eqnarray}
whose solutions have to be determined with periodic boundary conditions on the edges of the metamaterial unit cell and with vanishing spatial average.

Equations (\ref{eff-med}), (\ref{eff-tens}) and (\ref{eff-eqs}) are the main result of this paper and they show that the effective dielectric tensor $\varepsilon^{(eff)}$ and the tensors $\alpha^{(eff)}$ and $\beta^{(eff)}$ describing first and second order spatial dispersion can easily be evaluated after solving Eqs.(\ref{eff-eqs}) for a given periodic dielectric profile $\varepsilon(\R)$. In turn these equations share a common electrostatic structure and therefore they can be faced through a number of well known theoretical and numerical schemes. In addition we note that, to the best of our knowledge, the second and the third of Eqs.(\ref{eff-tens}) contain the simplest expressions for the tensors $\alpha^{(eff)}$ and $\beta^{(eff)}$ since they are easily obtained by averaging suitable mesoscopic fields. We conclude that, the discussed homogenization scheme provides a particularly simple way for predicting and analyzing, in addition to the local dielectric response, the spatial dispersion properties of an arbitrary tridimensional periodic dielectric medium in the long wavelength regime up to the second order.

\section{Electromagnetic chirality}
\subsection{Reciprocal bianisotropic metamaterial response}
Let us focus here on the situation where the contributions of order $\eta^2$ in Eq.(\ref{eff-med}) can be neglected. It is well known that in this case the effective medium first order nonlocal response is equivalent to a reciprocal bianisotropic response which is referred to as a electromagnetic chirality \cite{Serdyu}. In order to evaluate the chiral tensor let us note that Maxwell equations for the effective medium are left invariant by the transformation
\begin{eqnarray}
\overline{\textbf{E}'} &=& \overline{\textbf{E}}, \nonumber \\
\overline{\textbf{B}'} &=& \overline{\textbf{B}}, \nonumber \\
\overline{\textbf{D}'} &=& \overline{\textbf{D}} + \nabla \times \textbf{V}, \nonumber \\
\overline{\textbf{H}'} &=& \overline{\textbf{H}} - i \omega \textbf{V},
\end{eqnarray}
where $\textbf{V}$ is an arbitrary vector field, since Eqs. (\ref{macroscopic}), after defining $\overline{\textbf{B}} = \mu_0 \overline{\textbf{H}}$, can be rewritten as
\begin{eqnarray} \label{macroscopic-primed}
\nabla \times \overline{\textbf{E}'} &=& i \omega \overline{\textbf{B}'}, \nonumber \\
\nabla \times \overline{\textbf{H}'} &=& -i \omega \overline{\textbf{D}'}.
\end{eqnarray}
This shows at the same time that the fields $\textbf{D}$ and $\textbf{H}$ are not uniquely defined (as opposed to the fields $\textbf{E}$ and $\textbf{B}$) and that, for each possible vector field $\textbf{V}$, the constitutive relations
\begin{eqnarray} \label{constitutive}
\overline{\textbf{D}'} &=& \hat{\textbf{e}}_i \varepsilon_0 \left(
\varepsilon^{(eff)}_{ij} \overline{E'}_j +
\alpha^{(eff)}_{ijr} \frac{\partial \overline{E'}_j}{\partial x_r} \right) + \nabla \times \textbf{V}, \nonumber \\
\overline{\textbf{B}'} &=& \mu_0 \left( i\omega  \textbf{V} + \overline{\textbf{H}'} \right)
\end{eqnarray}
characterize the electromagnetic behavior of the effective medium. Choosing $\textbf{V} =\hat{\textbf{e}}_i \left( \frac{1}{\omega c\mu_0} \kappa_{ij} \overline{E}_j \right)$ for the arbitrary vector field $\textbf{V}$, where
\begin{equation} \label{kappa}
\kappa_{ij} = k_0 \left( -\frac{1}{2} \epsilon_{nmi} \alpha^{(eff)}_{nmj} + \frac{1}{4} \epsilon_{nmq} \alpha^{(eff)}_{nmq} \delta_{ij} \right)
\end{equation}
and $\epsilon_{ijk}$ is the Levi-Civita tensor, and exploiting the crucial antisymmetric property $\alpha^{(eff)}_{ijr} = \alpha^{(eff)}_{jir}$, Eqs.(\ref{constitutive}) become (see Appendix C)
\begin{eqnarray} \label{constitutive1}
\overline{\textbf{D}'} &=&  \varepsilon_0 \varepsilon'^{(eff)} \overline{\textbf{E}'} - \frac{i}{c} \kappa^T \overline{\textbf{H}'},  \nonumber \\
\overline{\textbf{B}'} &=& \frac{i}{c} \kappa \overline{\textbf{E}'} + \mu_0 \overline{\textbf{H}'}
\end{eqnarray}
where $\varepsilon'^{(eff)} = \varepsilon^{(eff)} + \kappa^T \kappa$. Equations (\ref{constitutive1}) shows that the effective medium has a reciprol magneto-electric coupling or electromagnetic chirality described by chiral tensor $\kappa$ of Eq.(\ref{kappa}). Inserting the second of Eqs.(\ref{eff-tens}) into Eq.(\ref{kappa}) we obtain
\begin{equation} \label{kappa1}
\kappa_{ij} = \eta k_0 \left[ \epsilon_{imj} \overline{\varepsilon f_m} +\left(\epsilon_{imn} \delta_{jq} + \frac{1}{2} \epsilon_{mqn} \delta_{ij} \right) \overline{\varepsilon f_m \partial_q f_n}\right].
\end{equation}
This relation is among the main results of the present paper and to the best of our knowledge it is the simplest expression for the effective medium chiral tensor since it is directly evaluated using the functions $f_i$ obtained by solving the first of Eqs.(\ref{eff-eqs}). Since we are dealing with a very general and tridimensional situation it is convenient to express the chirality tensor as \cite{Capoli}
\begin{equation}
\kappa = \frac{1}{3} Tr(\kappa) I + N + J
\end{equation}
where $Tr(\kappa) = \kappa_{ii}$ is the trace of the chirality tensor, $I$ is the identity tensor, $N$ is a symmetric trace-free tensor and $J$ is an antisymmetric tensor. Such decomposition allows to classify the reciprocal bianisotropic response \cite{Serdyu} of the effective medium and the three main classes are $Tr(\kappa) \neq 0, N \neq 0, J = 0$ for chiral media, $Tr(\kappa) = 0, N \neq 0, J = 0$ for pseudochiral media and $Tr(\kappa) = 0, N = 0, J \neq 0$ for omega media.

For the chirality tensor of Eq.(\ref{kappa1}) we obtain after some straightforward algebra
\begin{eqnarray}
Tr(\kappa) &=& \eta k_0 \left( \frac{1}{2} \epsilon_{mqn} \right) \overline{\varepsilon f_m \partial_q f_n}, \nonumber \\
N_{ij} &=& \eta k_0 \left[ \frac{1}{2} \left(\epsilon_{imn} \delta_{jq} + \epsilon_{jmn} \delta_{iq} \right) + \frac{1}{3} \epsilon_{mqn} \delta_{ij} \right] \overline{\varepsilon f_m \partial_q f_n}, \nonumber \\
J_{ij} &=& \eta k_0 \left[ \epsilon_{imj} \overline{\varepsilon f_m} + \frac{1}{2} \left(\epsilon_{imn} \delta_{jq} - \epsilon_{jmn} \delta_{iq} \right) \right] \overline{\varepsilon f_m \partial_q f_n}. \nonumber \\
\end{eqnarray}

\subsection{Spatial symmetries and chirality tensor structure}
A structure is chiral if it is distinguishable from its mirror image or, in other words, if its mirror image cannot be rigidly superposed onto it. If the structure is not chiral it is necessary invariant under a set of geometrical symmetries which affect the structure of the chirality tensor $\kappa$. In order to investigate the relation between geometrical and electromagnetic chirality it is necessary to identify the symmetries of the dielectric profile which either entail the vanishing or select a specific structure of the chiral tensor.

Let us suppose that the metamaterial is achiral or, in other words, that the mirror image of the dielectric profile through a plane orthogonal to the unit vector $\hat{\textbf{s}}$ can be rigidly superposed onto the original profile through the composition of a rotation of an angle $\theta$ around the unit vector $\hat{\textbf{r}}$ and a translation $\textbf{T}$. This implies that the dielectric permittivity is left invariant by the composition of these three transformations or
\begin{equation} \label{sym}
\epsilon(\R') = \epsilon(\R)
\end{equation}
where $\R'=X'_i \hat{\bf e}_i$  is the image of the point $\R=X_i \hat{\bf e}_i$ obtained through the spatial tranformation
\begin{equation} \label{transf}
X'_i = T_i + \tilde{R}_{ij}  X_j
\end{equation}
where $T_i$ are the components of the translation vector and $\tilde{R}_{ij} =  R_{ik} S_{kj}$ (or in matrix form $\tilde{R} = RS$) is the improper rotation obtained by composing the reflection $S$ and the rotation $R$ whose matrices are
\begin{eqnarray} \label{SR}
S_{nm} &=& \delta_{nm} - 2 s_n s_m \nonumber \\
R_{nm} &=& r_n r_m + (\delta_{nm} - r_n r_m) \cos \theta + \epsilon_{nkm} r_k \sin \theta. \nonumber \\
\end{eqnarray}
In Appendix D we fully investigate the way the symmetry of Eq.(\ref{sym}) affects electromagnetic chirality and as a result we obtain that such symmetry implies the vanishing of the chirality tensor apart from two specific situations where it imposes restrictions on the chirality tensor structure. With reference to Fig.1, these cases are: (a) the rotation is trivial $\theta=0$ or  the rotation unit vector is orthogonal to the reflection unit vector $\hat{\textbf{r}} \cdot \hat{\textbf{s}} = 0$; (b) the rotation and the reflection are such that $\sin^2 \left(\frac{\theta}{2}\right) ( \hat{\bf r}  \cdot \hat{\bf s})^2 = \frac{1}{2}$.

In order to discuss such symmetric situations where the chirality tensor does not wholly vanish it is convenient to consider the orthonormal basis
\begin{eqnarray} \label{uuuu}
\hat{\bf u}^{(1)} &=& \hat{\bf u}^{(2)} \times \hat{\bf u}^{(3)}, \nonumber \\
\hat{\bf u}^{(2)} &=& \frac{\hat{\bf u}^{(3)} \times \hat{\bf r}}{\sqrt{1-\left( \hat{\bf r} \cdot \hat{\bf u}^{(3)} \right)^2 }}
, \nonumber \\
\hat{\bf u}^{(3)} &=& \frac{- \sin \left(\frac{\theta}{2}\right) \hat{\bf s} \times \hat{\bf r} + \cos \left(\frac{\theta}{2}\right) \hat{\bf s}} {\sqrt{1-\sin^2 \left(\frac{\theta}{2}\right) ( \hat{\bf r}  \cdot \hat{\bf s})^2 }}.
\end{eqnarray}
Note that for $\theta = 0$ the rotation is the identity transformation so that the rotation vector $\hat{\bf r}$ is in this case an arbitrary unit vector.

In case (a), the dielectric profile is invariant under a pure reflection through a plane orthogonal to $\hat{\textbf{s}}$ or a suitable rotation around an axis belonging to the reflection plane is in order for restoring the dielectric profile after the reflection. As it is shown in Appendix D, in this case the unit vectors of Eqs.(\ref{uuuu}) are such that $\tilde{R} \hat{\bf u}^{(1)} = \hat{\bf u}^{(1)}$, $\tilde{R} \hat{\bf u}^{(2)} = \hat{\bf u}^{(2)}$ and $\tilde{R} \hat{\bf u}^{(3)} = - \hat{\bf u}^{(3)}$ so that the improper rotation $\tilde{R}$ is a pure reflection through the plane spanned by the unit vectors $\hat{\bf u}^{(1)}$ and $\hat{\bf u}^{(2)}$ and a rotation around $\hat{\bf u}^{(3)}$ of an angle $\psi = 0^\circ$. In other words case (a) deals with the situation where the medium admits a plane of reflection symmetry. In this case the structure of the chirality tensor allowed by this symmetry is (see Appendix D)
\begin{eqnarray} \label{kappa12}
\kappa = \eta \left( \kappa^{(13)} \hat{\bf u}^{(1)} \hat{\bf u}^{(3)T} + \kappa^{(31)} \hat{\bf u}^{(3)} \hat{\bf u}^{(1)T} \right. \nonumber \\
+ \left. \kappa^{(23)} \hat{\bf u}^{(2)} \hat{\bf u}^{(3)T} + \kappa^{(32)} \hat{\bf u}^{(3)} \hat{\bf u}^{(2)T} \right)
\end{eqnarray}
where $\kappa^{(13)}, \kappa^{(31)}, \kappa^{(23)}, \kappa^{(32)}$ are four independent complex scalars. From Eq.(\ref{kappa12}) it is evident that $Tr(\kappa)=0$ so that if the metamaterial admits a plane of reflection symmetry it is a pseudo-chiral-omega medium.

\begin{figure}
\center
\includegraphics*[width=0.48\textwidth]{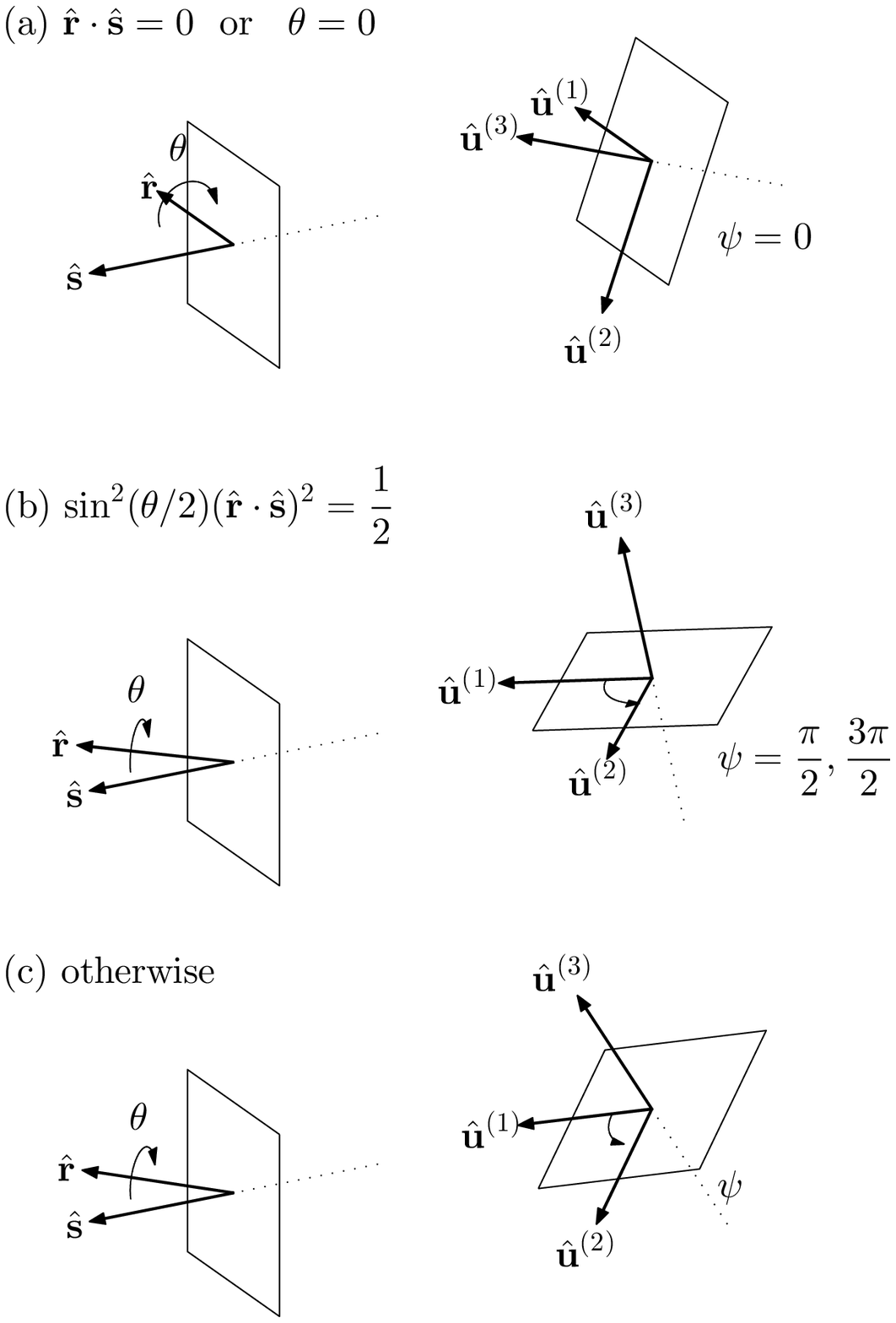}
\caption{(Color on-line) A reflection (dotted lines) through a plane orthogonal to the unit vector $\hat{\bf s}$ followed by a rotation (circular arrows) of an angle $\theta$ around the unit vector $\hat{\bf r}$ is equivalent to a reflection through the plane spanned by the unit vectors $\hat{\bf u}^{(1)}$ and $\hat{\bf u}^{(2)}$ and a rotation around the unit vector $\hat{\bf u}^{(3)}$ of an angle $\psi$ such that $\cos \psi = 1 -2 \sin^2 \left(\frac{\theta}{2}\right) ( \hat{\bf r}  \cdot \hat{\bf s})^2$. If the medium dielectric profile is invariant under the composition of the reflection and the rotation, three different situations are relevant. (a) The rotation is trivial $\theta=0$ or $\hat{\bf r}  \cdot \hat{\bf s} = 0$ so that $\psi = 0$. (b) The rotation and the reflection are such that $\sin^2 \left(\frac{\theta}{2}\right) ( \hat{\bf r}  \cdot \hat{\bf s})^2 = \frac{1}{2}$ so that $\psi= \frac{\pi}{2}$ or $\psi=\frac{3 \pi}{2}$. (c) Conditions (a) and (b) are not satisfied. In case (c) the chirality tensor vanishes whereas in cases (a) and (b) it has a reduced structure.}
\end{figure}

In case (b) the medium profile is restored, after the reflection, by a specific rotation in such a way that $\sin^2 \left(\frac{\theta}{2}\right) ( \hat{\bf r}  \cdot \hat{\bf s})^2 = \frac{1}{2}$. As it is shown in Appendix D, in this case the unit vectors of Eqs.(\ref{uuuu}) are such that $\tilde{R} \hat{\bf u}^{(1)} = \hat{\bf u}^{(2)}$, $\tilde{R} \hat{\bf u}^{(2)} = - \hat{\bf u}^{(1)}$ and $\tilde{R} \hat{\bf u}^{(3)} = - \hat{\bf u}^{(3)}$ so that the improper rotation $\tilde{R}$ is a reflection through the plane spanned by the unit vectors $\hat{\bf u}^{(1)}$ and $\hat{\bf u}^{(2)}$ and a rotation around the unit vector $\hat{\bf u}^{(3)}$ of an angle $\psi= 90^\circ$. In other words case (b) deals with the situation where the medium is left invariant (modulus a translation) by a reflection through a plane and a $90^\circ$ rotation around an axis orthogonal to such plane. In this case the structure of the chirality tensor allowed by this symmetry is
\begin{eqnarray} \label{kappasa}
\kappa &=& \eta \left[ \kappa^{(s)} \left( \hat{\bf u}^{(1)} \hat{\bf u}^{(2)T} + \hat{\bf u}^{(2)} \hat{\bf u}^{(1)T} \right) \right. \nonumber \\
       && \left. + \kappa^{(a)} \left( \hat{\bf u}^{(1)} \hat{\bf u}^{(1)T} - \hat{\bf u}^{(2)} \hat{\bf u}^{(2)T} \right) \right].
\end{eqnarray}
where $\kappa^{(s)}, \kappa^{(a)}$ are two independent complex scalars. From Eq.(\ref{kappasa}) it is evident that $Tr(\kappa)=0$ so that if the metamaterial has the symmetry of case (b) it is pseudo-chiral-omega medium.

We conclude that if the medium can be superposed to its mirror image the chirality tensor vanishes unless the rotation needed to restore the dielectric profile is around an axis orthogonal to the reflection plane and the rotation angle is either $0^\circ$ or $90^\circ$ and in both cases it is a pseudo-chiral-omega medium.

\subsection{Two-dimensional and one-dimensional electromagnetic chirality}
As an application of the above symmetry analysis let us consider two relevant situations where the medium is geometrically achiral and has a chiral electromagnetic response nonetheless.

The first situation is that of a medium which is invariant under translations along an axis, say the $Z-$axis, so that its underlying dielectric profile is $\varepsilon = \varepsilon(X,Y)$. This implies that the structure is left invariant by a reflection though the plane $Z=0$ followed by a rotation of zero angle $\theta = 0$ around any axis orthogonal to the $Z$-axis, so that we set $\hat{\bf s} = \hat{\bf e}_z$ and  $\hat{\bf r} = \hat{\bf e}_x$. Using such unit vectors, the basis vectors of Eqs.(\ref{uuuu}) becomes
\begin{eqnarray} \label{uuuu2D}
\hat{\bf u}^{(1)} &=& \hat{\bf e}_x , \nonumber \\
\hat{\bf u}^{(2)} &=& \hat{\bf e}_y , \nonumber \\
\hat{\bf u}^{(3)} &=& \hat{\bf e}_z.
\end{eqnarray}
so that since the medium symmetry is of the kind (a) (see the last paragraph), from Eqs.(\ref{kappa12}) we have
\begin{equation} \label{kappa2D}
\kappa =   \left(\begin{array}{ccc} 0 & 0 & \eta \kappa^{(13)} \\ 0 & 0 & \eta \kappa^{(23)} \\ \eta \kappa^{(31)} & \eta \kappa^{(32)} & 0 \end{array} \right)
\end{equation}
so that, in agreement with the considerations of the last paragraph, the metamaterial behaves as a pseudo-chiral-omega medium. Since such geometrically achiral medium is translationally invariant we conclude that Eq.(\ref{kappa2D}) contains the most general expression for the chirality tensor of a two-dimensional medium, i.e. it describes two-dimensional electromagnetic chirality. This situation also encompasses the case of a planar metamaterial \cite{Papako,Fedoto,Singhh,Zhangg,Yeeeee} (i.e. a very thin metamaterial sheet) since, after the reflection through a plane parallel to the medium plane, the sheet can be made to superpose the original sheet through a zero-angle rotation and a translation (which does not affect the chiral tensor structure) along the axis normal to the medium plane.

The second situation is that of a medium which is invariant under rotations around an axis, say the $Z-$axis, so that its underlying dielectric profile is $\varepsilon = \varepsilon(R,Z)$ where $R=\sqrt{X^2+Y^2}$. This implies that the structure is left invariant by a reflection though any plane containing the $Z$ axis followed by a rotation of zero angle $\theta = 0^\circ$ around the $Z$-axis. Therefore we set $\hat{\bf s} = \cos \phi \hat{\bf e}_x + \sin \phi \hat{\bf e}_y$, where $\phi$ is an arbitrary angle, and $\hat{\bf r} = \hat{\bf e}_z$. Using such unit vectors, the basis vectors of Eqs.(\ref{uuuu}) becomes
\begin{eqnarray} \label{uuuu1D}
\hat{\bf u}^{(1)} &=& \hat{\bf e}_z, \nonumber \\
\hat{\bf u}^{(2)} &=& \sin \phi \hat{\bf e}_x - \cos \phi \hat{\bf e}_y, \nonumber \\
\hat{\bf u}^{(3)} &=& \cos \phi \hat{\bf e}_x + \sin \phi \hat{\bf e}_y,
\end{eqnarray}
so that since the medium symmetry is of the kind (a) (see the last paragraph), from Eqs.(\ref{kappa12}) we have
\begin{widetext}
\begin{equation} \label{kappa1DD}
\kappa =  \eta \left(\begin{array}{ccc}
(\kappa^{(23)}+\kappa^{(32)}) \sin \phi \cos \phi  &  \kappa^{(23)} \sin^2 \phi - \kappa^{(32)} \cos^2 \phi  & \kappa^{(31)} \cos \phi \\
-\kappa^{(23)} \cos^2 \phi + \kappa^{(32)} \sin^2 \phi  &  -(\kappa^{(23)}+\kappa^{(32)}) \sin \phi \cos \phi  & \kappa^{(31)} \sin \phi \\
\kappa^{(13)} \cos \phi & \kappa^{(13)} \sin \phi & 0
\end{array} \right)
\end{equation}
\end{widetext}
where $\kappa^{(13)}(\phi)$, $\kappa^{(31)}(\phi)$, $\kappa^{(23)}(\phi)$ and $\kappa^{(32)}(\phi)$ are functions of $\phi$. The entries of the chirality tensor of Eq.(\ref{kappa1DD}) cannot depend on the arbitrary angle $\phi$ so that imposing that each $\kappa_{ij}$ is constant we get $\kappa^{(13)}=\kappa^{(31)}=0$ and $\kappa^{(23)} = - \kappa^{(32)} = \kappa_0$ and therefore Eq.(\ref{kappa1DD}) yields
\begin{equation} \label{kappa1D}
\kappa =   \left(\begin{array}{ccc} 0 & \eta \kappa_0 & 0 \\ -\eta \kappa_0  &  0  & 0 \\ 0 & 0 & 0 \end{array} \right).
\end{equation}
Therefore, a rotationally invariant metamaterial around the $Z$-axis behaves as an omega-medium whose chirality tensor depends on a single parameter $\kappa_0$. This situation encompasses the case of a one-dimensional metamaterial for which $\epsilon= \epsilon(Z)$ so that we conclude that  Eq.(\ref{kappa1D}) contains the most general expression for the chirality tensor of a one-dimensional metamaterial, i.e. it describes one-dimensional electromagnetic chirality.

\section{Electromagnetic chirality of one-dimensional metamaterials}
The description of the effective medium response, i.e. the evaluation of the tensors of Eqs.(\ref{eff-tens}), is achieved by solving Eqs.(\ref{eff-eqs}) for a specific dielectric distribution and numerical integration is usually unavoidable. However there is a situation where electromagnetic chirality can be fully discussed without resorting to numerical analysis and it is the case of one-dimensional metamaterials.

Let us consider a metamaterial whose dielectric profile is such that $\varepsilon_r(z)=\varepsilon_r(z+d)$, i.e. it is a periodic function of the single cartesian coordinate $z$, whose period $d$ is such that $d \ll \lambda$ for the homogenization theory to be applicable. After introducing the fast spatial coordinates $(X,Y,Z)=(x,y,z)/\eta$ where $\eta = d / \lambda$ is the homogenization parameter, the dielectric profile $\varepsilon(Z) = \varepsilon_r(\eta Z)$ of Eq.(\ref{dielectric}) is a periodic function of period $\lambda$. In Appendix E we show that for such dielectric distribution $\varepsilon(Z)$, the first of Eqs.(\ref{eff-eqs}) can be analytically solved so that the effective permittivity of the first of Eqs.(\ref{eff-tens}) and the chirality tensor of Eq.(\ref{kappa1}) can be analytically evaluated and they are
\begin{equation} \label{1D-eff}
\varepsilon^{(eff)} = \left(\begin{array}{ccc} \overline{\varepsilon} & 0 & 0 \\ 0  &  \overline{\varepsilon}  & 0 \\ 0 & 0 & \left[\overline{\varepsilon^{-1}}\right]^{-1} \end{array} \right), \hspace{0.5cm}
\kappa = \left(\begin{array}{ccc} 0 & \eta \kappa_0 & 0 \\ -\eta \kappa_0  &  0  & 0 \\ 0 & 0 & 0 \end{array} \right).
\end{equation}
where
\begin{eqnarray} \label{1D-kappa0-body}
\kappa_0 &=& \left[\overline{\varepsilon^{-1}}\right]^{-1} \frac{2 \pi }{ \lambda^2} \int_0^\lambda dZ_1 \int_0^\lambda dZ_2 \frac{\varepsilon(Z_1)}{\varepsilon(Z_2)} \left[ \left(\frac{Z_1-Z_2}{\lambda}\right)  \right. \nonumber \\ &&-
\left. \frac{1}{2} \textrm{sign} \left(\frac{Z_1-Z_2}{\lambda}\right) \right].
\end{eqnarray}
The effective dielectric tensor in the first of Eqs.(\ref{1D-eff}) coincides with the well-known result of the zeroth-order homegenization theory of layered media. The structure of the chirality tensor in the second of Eqs.(\ref{1D-eff}) shows that the effective medium behaves as an omega-medium and this agrees with the discussion of the last paragraph, see Eq.(\ref{kappa1D}), since the considered one-dimensional medium is in particular rotationally invariant around the $Z$-axis. It is also remarkable that the parameter $\kappa_0$ of Eq.(\ref{1D-kappa0-body}) governing electromagnetic chirality in the present case can be directly evaluated from the dielectric profile by quadratures.

\section{Conclusions}
To sum up we have investigated the electromagnetic response of a metamaterial with non-magnetic inclusions by a multiscale analysis and we have included spatial nonlocality up to the second order. The resulting description is very simple and compact since the three tensors characterizing the effective medium response are shown to be averages (on the unit cell) of suitable fast-varying fields which, in turn, are obtained from the dielectric distribution by solving a small number of electrostatic equations. If second order nonlocal response can be neglected, the medium response can be recast into the standard bianisotropic form and this has allowed us to obtain a compact expression for the chirality tensor of the effective medium. We have exploited our approach to investigate the relation between geometrical and electromagnetic chirality by proving that the latter is shown, in addition to chiral media, even by an achiral medium whose mirror image can be superposed onto it by means of a $0^\circ$ or $90^\circ$ rotation around an axis orthogonal to the reflection plane. We have deduced, as specific relevant examples, the chirality tensor structure of two-dimensional and one-dimensional media and, for the formers, we have obtained a closed form expression for the chirality parameter which can be evaluated from the inclusions dielectric profile by means of quadratures.

\section*{ACKNOWLEDGEMENT}
The authors thank the U.S. Army International Technology Center Atlantic for financial support (Grant No. W911NF-14-1-0315).

\appendix
\section{The electrostatic equation}
Let us consider the equation
\begin{equation} \label{fund}
\nabla_\R \cdot \left( \varepsilon \nabla_\R \Psi \right) = \Phi
\end{equation}
whose source term $\Phi(\R)$ has the metamaterial periodicity with vanishing spatial average $\overline{\Phi} = 0$ and whose solution $\Psi(\R)$ is required to have the metamaterial periodicity. Equation (\ref{fund}) can be written as $\hat{L} \Psi = \Phi$ where the operator $\hat{L}$ is
\begin{equation} \label{operator}
\hat{L} = \nabla_\R \cdot \left( \varepsilon \nabla_\R  \right),
\end{equation}
so that solving Eq.(\ref{fund}) amounts to the problem of inverting the operator $\hat{L}$ on the space of functions having the metamaterial periodicity.

Note that the relation $\nabla_\R \cdot \left(  \varepsilon \Psi^*_0 \nabla_\R \Psi_0 \right) = \varepsilon \left| \nabla_\R \Psi_0 \right|^2 + \Psi^*_0 \hat{L} \Psi$ holds for any function $\Psi_0$ and therefore, if $\Psi_0$ satisfies the homogeneous equation $\hat{L} \Psi_0 = 0$ and it has the metamaterial periodicity we have
\begin{eqnarray}
\int_C d^3\R \varepsilon \left| \nabla_\R \Psi_0 \right|^2 &=& \int_C d^3\R \nabla_\R \cdot \left(  \varepsilon \Psi^*_0 \nabla_\R \Psi_0 \right) \nonumber \\
&=& \int_{\partial C} dS \varepsilon \Psi^*_0 \hat{\bf{n}} \cdot \nabla_\R \Psi_0 = 0
\end{eqnarray}
where Green theorem has been used together with the periodicity of both $\varepsilon$ and $\Psi_0$. The real and imaginary parts of this equation read
\begin{eqnarray}
\int_C d^3\R \textrm{Re}(\varepsilon) \left| \nabla_\R \Psi_0 \right|^2 &=& 0 \nonumber \\
\int_C d^3\R \textrm{Im}(\varepsilon) \left| \nabla_\R \Psi_0 \right|^2 &=& 0
\end{eqnarray}
and therefore, since we are considering dielectrics without gain for which $\textrm{Im}(\varepsilon) > 0$, we obtain $\nabla_\R \Psi_0 = \textbf{0}$ or $\Psi_0 = const.$. This implies that the function $\Psi' = \Psi + \Psi_0$ satisfies Eq.(\ref{fund}) and the boundary conditions if $\Psi(\R)$ does and therefore Eq.(\ref{fund}) and the periodic boundary conditions determines $\Psi$ up to an arbitrary constant.

Setting $\varepsilon = 1$ in Eq.(\ref{fund}) we obtain the Poisson equation $\nabla_\R^2 \Psi  = \Phi$ for which the above considerations holds so that $\Psi$ has to be determined with the metamaterial periodicity up to an arbitrary constant.

\section{Tensor properties of the effective medium response}
A straightforward calculation shows that the relation
\begin{equation}
\Psi \hat{L} \Phi - \Phi \hat{L} \Psi = \nabla_\R \cdot \left[ \varepsilon \left(\Psi \nabla_\R \Phi - \Phi \nabla_\R \Psi \right) \right],
\end{equation}
where $\hat{L}$ is the operator defined in Eq.(\ref{operator}), holds for any pair of functions $\Psi$ and $\Phi$ so that, if both these functions have the metamaterial periodicity after using Green theorem we obtain
\begin{equation} \label{ident}
\int_C d^3 \R \Psi \hat{L} \Phi = \int_C d^3 \R  \Phi \hat{L} \Psi.
\end{equation}
Let us consider a function $\Psi(\R)$ which has the metamaterial periodicity. The average $\overline{\varepsilon \partial_i \Psi}$ can be rewritten as
\begin{equation}
\overline{\varepsilon \partial_i \Psi} = \frac{1}{V} \int_C d^3 \R \varepsilon \partial_i \Psi = - \frac{1}{V} \int_C d^3 \R   \Psi \partial_i \varepsilon
\end{equation}
where integration by parts and the periodicity of both $\varepsilon$ and $\Psi$ have been used. Using Eq.(\ref{f0}) we obtain
\begin{equation}
\overline{\varepsilon \partial_i \Psi} = \frac{1}{V} \int_C d^3 \R  \Psi  \hat{L} f_i
\end{equation}
so that Eq.(\ref{ident}) with $\Phi = f_i$ yields
\begin{equation} \label{basic}
\overline{\varepsilon \partial_i \Psi} = \frac{1}{V} \int_C d^3 \R  f_i \hat{L} \Psi .
\end{equation}
Using Eq.(\ref{basic}) with $\Psi = f_j$, the rank-two tensor $\Delta \varepsilon^{(eff)}_{ij} = \overline{ \varepsilon \partial_i f_j}$ can be written as
\begin{equation}
\Delta \varepsilon^{(eff)}_{ij} =
\frac{1}{V} \int_C d^3 \R  f_i \hat{L} f_j  =
\frac{1}{V} \int_C d^3 \R  f_j \hat{L} f_i
\end{equation}
where Eq.(\ref{ident}) has been used and therefore $\Delta \varepsilon^{(eff)}_{ij} = \Delta \varepsilon^{(eff)}_{ji}$.

Setting $\Psi = W_{rj}$, Eq.(\ref{basic}) becomes
\begin{eqnarray}
\overline{\varepsilon \partial_i W_{rj}} &=&
\frac{1}{V} \int_C d^3 \R  f_i \hat{L} W_{rj}  \nonumber \\ &=&
- \frac{1}{V} \int_C d^3 \R  f_i \left[ \partial_r \left( \varepsilon f_j\right) + Q_{rj}  \right]
\end{eqnarray}
where Eqs.(\ref{f1}) and (\ref{Q}) have been used together with the property $\overline{f_i} = 0$. After integrating by parts the first term we get
\begin{eqnarray}
\overline{\varepsilon \partial_i W_{rj}} = \overline{  \varepsilon\left( \partial_r f_i \right) f_j  - Q_{rj} f_i }.
\end{eqnarray}
Therefore the rank-three tensor $\alpha^{(eff)}_{ijr} = \eta \overline{ \varepsilon \left(  \delta_{ir} f_j + \partial_i W_{rj} \right)}$ becomes
\begin{eqnarray}
\alpha^{(eff)}_{ijr} &=& \eta \overline{ \left( Q_{ri} f_j - Q_{rj} f_i \right) }
\end{eqnarray}
where use has been made of the fact that $\overline{f_i} = \overline{f_j} = 0$.

Setting $\Psi = v_j$ Eq.(\ref{basic}) becomes
\begin{eqnarray}
\overline{\varepsilon \partial_i v_j} &=&
\frac{1}{V} \int_C d^3 \R  f_i \hat{L} v_j =
\frac{1}{V} \int_C d^3 \R A_{rj} \varepsilon \partial_r f_i \nonumber \\
\end{eqnarray}
where the second of Eqs.(\ref{R}) has been used and integration by parts has been performed. Noting that Eq.(\ref{Q}) and the second of Eqs.(\ref{magneto}) can be combined to yield $\varepsilon\partial_r f_i = - \varepsilon \delta_{ri} - \frac{1}{k_0^2} \nabla_\R^2 A_{ri} + \overline{Q_{ri}}$ we obtain
\begin{eqnarray}
\overline{\varepsilon \partial_i v_j} &=&
\frac{1}{V} \int_C d^3 \R A_{rj} \left( - \varepsilon \delta_{ri} - \frac{1}{k_0^2} \nabla_\R^2 A_{ri}  \right)  \nonumber \\ &=&
- \overline{\varepsilon A_{ij} }  + \frac{1}{k_0^2} \overline{ \nabla_\R A_{ri} \cdot  \nabla_\R A_{rj} }
\end{eqnarray}
where the property $\overline{A_{rj}} = 0$ has been used and integration by parts has been performed. Therefore, for the rank-two tensor $\gamma^{(eff)}_{ij} = \overline{\varepsilon \left(A_{ij}+ \partial_i v_j\right)}$ we have
\begin{equation}
\gamma^{(eff)}_{ij} = \frac{1}{k_0^2} \overline{ ( \partial_s A_{ri})( \partial_s A_{rj}) }.
\end{equation}
Setting $\Psi = R_{rsj}$, Eq.(\ref{basic}) becomes
\begin{eqnarray} \label{second}
\overline{\varepsilon \partial_i R_{rsj}} &=&
\frac{1}{V} \int_C d^3 \R  f_i \hat{L} R_{rsj} =
\frac{1}{V} \int_C d^3 \R   \left[  W_{sj} \left ( \varepsilon \partial_r f_i \right)  \right]
 \nonumber \\ &&
- \frac{1}{V} \int_C d^3 \R   \left[  \varepsilon \left( f_i \partial_r W_{sj} + \delta_{rs} f_i f_j \right) \right]
\end{eqnarray}
where the first of Eqs.(\ref{R}) and Eq.(\ref{P}) have been used together with the property $\overline{f_i} = 0$ and an integration by parts has been performed. From Eq.(\ref{f1}) we have $\varepsilon \partial_r f_i = -\varepsilon \delta_{ri}- \partial_r (\varepsilon f_i) + \overline{Q_{ri}} - \hat{L} W_{ri}$ which inserted in Eq.(\ref{second}), after using the property $\overline{W_{sj}}=0$ and integrating by parts, yields
\begin{eqnarray}
\overline{\varepsilon \partial_i R_{rsj}} &=&
\frac{1}{V} \int_C d^3 \R  \left[\varepsilon \left( \delta_{rs} f_i f_j + \delta_{ri} W_{sj} \right) - W_{sj} \hat{L} W_{ri} \right]. \nonumber \\ &&
\end{eqnarray}
With the help of this equation, the fourth-rank tensor $\beta^{(eff)}_{ijsr} = \frac{\eta^2}{2} \overline{\varepsilon \left(  \delta_{is} W_{rj} + \delta_{ir} W_{sj} + \partial_i R_{srj} + \partial_i R_{rsj} \right)}$ becomes
\begin{eqnarray}
\beta^{(eff)}_{ijsr} &=& - \frac{\eta^2}{V} \int_C d^3 \R \left( \varepsilon \delta_{rs} f_i f_j \right) \nonumber \\ && -
\frac{\eta^2}{2V} \int_C d^3 \R  \left( W_{sj} \hat{L} W_{ri} + W_{rj} \hat{L} W_{si} \right). \nonumber \\
\end{eqnarray}
Using Eq.(\ref{ident}) it is straightforward to show that $\beta^{(eff)}_{ijsr} = \beta^{(eff)}_{jisr}$, i.e. the tensor is symmetric with respect to the pair of indices $ij$. After symmetrizing these indices, i.e. setting $\beta^{(eff)}_{ijsr} = \frac{1}{2} \left(\beta^{(eff)}_{ijsr} + \beta^{(eff)}_{jisr}\right)$, and using Eq.(\ref{f1}) we obtain
\begin{eqnarray}
\beta^{(eff)}_{ijsr} &=&  \eta^2 \overline{\varepsilon \left( -\delta_{rs} f_i f_j \right)} \nonumber \\ && +
\frac{\eta^2}{4}\overline{ \left( Q_{ri} W_{sj} + Q_{si} W_{rj} + Q_{rj} W_{si} + Q_{sj} W_{ri} \right) } \nonumber \\ && -
\frac{\eta^2}{4}\overline{ \varepsilon  \left( f_i \partial_r W_{sj} +  f_i \partial_s W_{rj} +  f_j \partial_r W_{si} +  f_j \partial_s W_{ri} \right)}. \nonumber \\
\end{eqnarray}
or, equivalently
\begin{eqnarray}
\beta^{(eff)}_{ijsr} &=&
\frac{\eta^2}{4}\overline{ \left( Q_{ri} W_{sj} + Q_{si} W_{rj} + Q_{rj} W_{si} + Q_{sj} W_{ri} \right) } \nonumber \\ && -
\frac{\eta^2}{4}\overline{ \left[ f_i \left(P_{rsj} + P_{srj}\right) + f_j \left( P_{rsi} + P_{sri} \right) \right]}. \nonumber \\
\end{eqnarray}

\section{Effective medium electromagnetic chirality}

In order to show that Eqs.(\ref{constitutive1}) hold, let us first prove that the relation
\begin{equation} \label{alf-kappa}
\alpha^{(eff)}_{ijr} + \frac{1}{k_0} \epsilon_{irk} \kappa_{kj}   =  \frac{1}{k_0} \epsilon_{jrk} \kappa_{ki}
\end{equation}
is satisfied. Using Eq.(\ref{kappa}) we have
\begin{eqnarray}
&& \alpha^{(eff)}_{ijr} + \frac{1}{k_0} \left(\epsilon_{irk} \kappa_{kj} - \epsilon_{jrk} \kappa_{ki} \right) = \alpha^{(eff)}_{ijr} \nonumber \\
&& +\frac{1}{2} \left( - \epsilon_{irk} \epsilon_{nmk} \alpha^{(eff)}_{nmj} + \epsilon_{jrk} \epsilon_{nmk} \alpha^{(eff)}_{nmi} \right) \nonumber \\
&& - \frac{1}{2} \epsilon_{nmq} \alpha^{(eff)}_{nmq}  \epsilon_{ijr}
\end{eqnarray}
which, after exploiting the relation $\epsilon_{abk}\epsilon_{cdk} = \delta_{ac} \delta_{bd} - \delta_{ad} \delta_{bc}$ and the antisymmetric property $\alpha^{(eff)}_{ijr} = \alpha^{(eff)}_{jir}$, becomes
\begin{eqnarray} \label{proof}
&& \alpha^{(eff)}_{ijr} + \frac{1}{k_0} \left(\epsilon_{irk} \kappa_{kj} - \epsilon_{jrk} \kappa_{ki} \right) = \nonumber \\ && \alpha^{(eff)}_{ijr} + \alpha^{(eff)}_{rij} + \alpha^{(eff)}_{jri} - \frac{1}{2} \epsilon_{nmq} \alpha^{(eff)}_{nmq}  \epsilon_{ijr}.
\end{eqnarray}
The tensor $F_{ijr} = \alpha^{(eff)}_{ijr} + \alpha^{(eff)}_{rij} + \alpha^{(eff)}_{jri}$ is completely antisymmetric since, using again the antisymmetric property $\alpha^{(eff)}_{ijr} = \alpha^{(eff)}_{jir}$, it is straighforward proving that $F_{ijr} = -F_{jir} = - F_{irj} = - F_{rji}$. Since every completely antisymmetric three-rank tensor in a tridimensional space is proportional to the Levi-Civita tensor, the relation $F_{ijr} = F \epsilon_{ijr}$ holds and the scalar $F$ is given by $F= \frac{1}{6} \epsilon_{nmq} F_{nmq}$ since the relation $\epsilon_{nmq} \epsilon_{nmq} = 6$ holds. Using the antisymmetric property $\alpha^{(eff)}_{ijr} = \alpha^{(eff)}_{jir}$ we obtain $F = 3 \epsilon_{nmq} \alpha^{(eff)}_{nmq}$ so that
\begin{equation}
\alpha^{(eff)}_{ijr} + \alpha^{(eff)}_{rij} + \alpha^{(eff)}_{jri} = F_{ijr} = \frac{1}{2} \epsilon_{nmq} \alpha^{(eff)}_{nmq} \epsilon_{ijr}
\end{equation}
which, inserted into Eq.(\ref{proof}), shows that Eq.(\ref{alf-kappa}) is satisfied.

Inserting the expression $\textbf{V} = \hat{\textbf{e}}_i \left( \frac{1}{\omega c \mu_0}  \kappa_{ij} \overline{E}_j \right)$ into the second of Eqs.(\ref{constitutive}) and using the relation $\overline{\textbf{E}'} = \overline{\textbf{E}}$ we obtain
\begin{equation} \label{B'}
\overline{\textbf{B}'} = \hat{\textbf{e}}_i \left( \frac{i}{c} \kappa_{ij} \overline{E'}_j \right) + \mu_0 \overline{\textbf{H}'}
\end{equation}
which coincides with the second of Eqs.(\ref{constitutive1}). Inserting the expression $\textbf{V} = \hat{\textbf{e}}_i \left( \frac{1}{\omega c \mu_0}  \kappa_{ij} \overline{E}_j \right)$ into the first of Eqs.(\ref{constitutive}) and using the relation $\overline{\textbf{E}'} = \overline{\textbf{E}}$ we obtain
\begin{eqnarray}
\overline{\textbf{D}'} &=& \hat{\textbf{e}}_i \varepsilon_0 \left[
\varepsilon^{(eff)}_{ij} \overline{E'}_j +
\left( \alpha^{(eff)}_{ijr} + \frac{1}{k_0} \epsilon_{irk} \kappa_{kj} \right) \frac{\partial \overline{E'}_j}{\partial x_r} \right] \nonumber \\ &=&
\hat{\textbf{e}}_i \varepsilon_0 \left[ \varepsilon^{(eff)}_{ij} \overline{E'}_j - \frac{1}{k_0} \kappa_{ki} \epsilon_{krj} \frac{\partial \overline{E'}_j}{\partial x_r} \right]
\end{eqnarray}
where Eq.(\ref{alf-kappa}) has been used together with the antisymmetry of the Levi-Civita tensor. Exploiting the first of Eqs.(\ref{macroscopic-primed}) this equation becomes
\begin{eqnarray}
\overline{\textbf{D}'} &=&
\hat{\textbf{e}}_i \varepsilon_0 \left( \varepsilon^{(eff)}_{ij} \overline{E'}_j - i c \kappa_{ki} \overline{B'}_k \right)
\end{eqnarray}
which, using Eq.(\ref{B'}), yields
\begin{eqnarray}
\overline{\textbf{D}'} &=&
\hat{\textbf{e}}_i \left[ \varepsilon_0 \left(\varepsilon^{(eff)}_{ij} + \kappa_{ki} \kappa_{kj} \right) \overline{E'}_j - \frac{i}{c} \kappa_{ki} \overline{H'}_k \right]
\end{eqnarray}
which in turn coincides with the first of Eqs.(\ref{constitutive1}).

\section{Structure of the chirality tensor imposed by symmetry}
In order to discuss the impact of the symmetry of the dielectric profile on the chirality tensor, it is essential to investigate preliminary the transformation rule of the fields $f_i$ induced by the dielectric symmetry. Differentiating the permittivity $\epsilon(\R')$ with respect $X_i'$ and using Eq.(\ref{sym}) we obtain
\begin{equation} \label{par-ep}
(\partial_i \epsilon)_{\R'} = \tilde{R}_{ij} (\partial_j \epsilon)_{\R}.
\end{equation}
where $\tilde{R}_{ij} = R_{ik} S_{kj}$. Let us now consider the functions
\begin{equation} \label{fp}
f'_i (\R) = f_i (\R')
\end{equation}
for which, using again Eq.(\ref{sym}), we have
\begin{eqnarray} \label{inv-oper}
\left[ \nabla_\R \cdot \left(\epsilon \nabla_\R f'_i \right) \right]_{\R} &=&
\frac{\partial}{\partial X_n} \left[ \epsilon(\R') \frac{\partial f_i (\R')}{\partial X_n} \right] \nonumber \\ &=&
\tilde{R}_{rn} \tilde{R}_{sn} \left[ \partial_r \left(\epsilon \partial_s f_i \right) \right]_{\R'} \nonumber \\ &=&
\left[ \nabla_\R \cdot \left(\epsilon \nabla_\R f_i \right) \right]_{\R'}
\end{eqnarray}
where the relation $\tilde{R}_{rn} \tilde{R}_{sn} = \delta_{rs}$, stating the orthogonality of the improper rotation $\tilde{R}$, have been used. Combining Eq.(\ref{inv-oper}), the first of Eqs.(\ref{eff-eqs}) evaluated at $\R'$ and Eq.(\ref{par-ep}) we obtain
\begin{equation}
\nabla_\R \cdot \left[ \epsilon \nabla_\R \left( \tilde{R}_{ij} f'_i \right) \right]  = -\partial_j \epsilon
\end{equation}
which structurally coincides with the first of Eqs.(\ref{eff-eqs}) and which shows, by the uniquity of its solution, that $\tilde{R}_{ij}  f'_i = f_j$. From Eq.(\ref{fp}), we thus conclude that the symmetry of the dielectric permittivity entails for the functions $f_i$ the property
\begin{equation} \label{fpp}
f_i(\R') = \tilde{R}_{ij} f_j(\R).
\end{equation}
From Eq.(\ref{kappa1}) we note that the chirality tensor depends on the rank-one tensor $\overline{\varepsilon f_m}$ and rank-three tensor $\overline{\varepsilon f_m \partial_q f_n}$ whose geometrical properties have to be separately considered. Using this equation, exploting the fact that the improper rotation is orthogonal (so that $|det(R)|=1$) and noting that the transformation in Eqs.(\ref{transf}) superposes a metamaterial unit cell $C$ onto another metamaterial unit cell $C'$ we have
\begin{eqnarray} \label{vec}
\overline{\varepsilon f_m} &=& \frac{1}{V} \int_{C'} d^3\R' \varepsilon(\R') f_m (\R') \nonumber \\ &=&
 \frac{1}{V} \int_{C} d^3\R \varepsilon(\R) \tilde{R}_{mi} f_i (\R) = \tilde{R}_{mi} \overline{\varepsilon f_i}.
\end{eqnarray}
After differentiation, Eq.(\ref{fpp}) yields
\begin{equation}
\left(\partial_j f_i\right)_{\R'} = \tilde{R}_{jn} \tilde{R}_{im} \left(\partial_n f_m \right)_{\R}
\end{equation}
so that, reasoning as in Eq.(\ref{vec}) we have
\begin{eqnarray} \label{tens}
\overline{\varepsilon f_m \partial_q f_n} &=& \frac{1}{V} \int_{C'} d^3\R' \varepsilon(\R') f_m (\R') (\partial_q f_n)_{\R'}
\nonumber \\ &=&
\frac{1}{V} \int_{C} d^3\R \varepsilon(\R) \tilde{R}_{mi} f_i(\R) \tilde{R}_{qp} \tilde{R}_{nj} (\partial_p f_j)_{\R'}
\nonumber \\ &=& \tilde{R}_{mi} \tilde{R}_{qp} \tilde{R}_{nj} \overline{\varepsilon f_i \partial_p f_j}.
\end{eqnarray}
Equations (\ref{vec}) and (\ref{tens}) state that the vector $\overline{\varepsilon f_m}$ and the tensor $\overline{\varepsilon f_m \partial_q f_n}$ are geometrical invariants of the improper rotation $\tilde{R}$.

Since $\tilde{R}$ is an improper rotation ($\tilde{R}\tilde{R}^T = I$ and $det(\tilde{R}) = -1$), it is diagonalizable and its eigenvalues are $\lambda_1=e^{i\psi}$, $\lambda_2 =e^{-i\psi}$, $\lambda_3=-1$ (where $0\leq \psi < 2 \pi$). In order to relate $\psi$ to the original reflection and rotation we note that  $Tr(\tilde{R}) = \lambda_1 + \lambda_2 + \lambda_3 =  2 \cos \psi -1$ whereas from Eqs.(\ref{SR}) we obtain $Tr(\tilde{R}) = R_{ik} S_{ki} = 1-4 \sin^2 \left(\frac{\theta}{2}\right) (r_i s_i)^2$ so that, equating these two expressions, we get
\begin{equation} \label{psi}
\cos \psi = 1 -2 \sin^2 \left(\frac{\theta}{2}\right) ( \hat{\bf r}  \cdot \hat{\bf s})^2.
\end{equation}
The matrix $\tilde{R}$ admits three eigenvectors $\hat{\bf w}^{(1)},\hat{\bf w}^{(2)},\hat{\bf w}^{(3)}$, such that $\tilde{R} \hat{\bf w}^{(n)} = \lambda_n \hat{\bf w}^{(n)}$, and they satisfy the orthogonality and completeness relations
\begin{eqnarray} \label{MM}
w_i^{(n)*} w_i^{(m)} &=& \delta_{nm}, \nonumber \\
w_i^{(n)*} w_j^{(n)} &=& \delta_{ij}.
\end{eqnarray}
Therefore the matrix $\tilde{R}$ admits the spectral decomposition $\tilde{R}_{ij}= \lambda_n w_i^{(n)} w_j^{(n)*}$ so that, after setting
\begin{eqnarray} \label{vec-tens}
\overline{\varepsilon f_i} &=& V_n w_i^{(n)}, \nonumber \\
\overline{\varepsilon f_i \partial_p f_j} &=& W_{mqn} w_i^{(m)} w_p^{(q)} w_j^{(n)},
\end{eqnarray}
with the help Eqs.(\ref{MM}), Eqs.(\ref{vec}) and (\ref{tens}) become
\begin{eqnarray} \label{spectr}
V_n &=& \lambda_n V_n, \nonumber \\
W_{mqp} &=& \lambda_m \lambda_q \lambda_p W_{mqp}
\end{eqnarray}
where no summation is performed over the indices.

From Eqs.(\ref{spectr}) it is evident that all the components of $V_n$ and $W_{mqp}$ vanish unless the coefficients $\lambda_n$ and $\lambda_m \lambda_q \lambda_p$ are equal to $1$. Since the allowed values for $\lambda_n$ are $\left\{-1, e^{i\psi}, e^{-i\psi}\right\}$ and those for $\lambda_m \lambda_q \lambda_p$ are $\left\{-1, e^{i\psi}, e^{-i\psi}, -e^{2i\psi}, -e^{-2i\psi}, e^{3i\psi}, -e^{3i\psi} \right\}$, it is evident that the only possible cases where some of the components of $V_n$ and $W_{mqp}$ do not vanish are for $\psi = \left\{ 0, \frac{\pi}{2}, \frac{3 \pi}{2}, \frac{2\pi}{3}, \frac{4\pi}{3} \right\}$. As a consequence these values of $\psi$, corresponding to classes of original reflections and rotations, leads to nonvanishing chiral tensors.

Let us now discuss the various classes of original reflections and rotations corresponding to the allowed values of $\psi$. In order to simplify the treatment let us consider the three vectors
\begin{eqnarray} \label{uuu}
\hat{\bf u}^{(1)} &=& \hat{\bf u}^{(2)} \times \hat{\bf u}^{(3)}, \nonumber \\
\hat{\bf u}^{(2)} &=& \frac{\hat{\bf u}^{(3)} \times \hat{\bf r}}{\sqrt{1-\left( \hat{\bf r} \cdot \hat{\bf u}^{(3)} \right)^2 }}
, \nonumber \\
\hat{\bf u}^{(3)} &=& \frac{- \sin \left(\frac{\theta}{2}\right) \hat{\bf s} \times \hat{\bf r} + \cos \left(\frac{\theta}{2}\right) \hat{\bf s}} {\sqrt{1-\sin^2 \left(\frac{\theta}{2}\right) ( \hat{\bf r}  \cdot \hat{\bf s})^2 }}
\end{eqnarray}
which are easily seen to have real components and to form an orthonormal and positively oriented basis, i.e.
\begin{eqnarray} \label{vers1}
\hat{\bf u}^{(n)} \cdot \hat{\bf u}^{(m)} &=& \delta_{nm}, \nonumber \\
\hat{\bf u}^{(n)} \times \hat{\bf u}^{(m)} &=& \epsilon_{nmr} \hat{\bf u}^{(r)}.
\end{eqnarray}
The second of Eqs.(\ref{vers1}) implies that the vectors of Eqs.(\ref{uuu}) satisfy the relation
\begin{equation} \label{vers2}
\epsilon_{imj} u_m^{(k)} = \frac{1}{2} \epsilon_{nkm} \left( u_i^{(n)} u_j^{(m)} - u_i^{(m)} u_j^{(n)} \right).
\end{equation}
which is particularly useful for the developing of the present analysis.

{\bf Case} $\psi=0$. From Eq.(\ref{psi}) we deduce that this case geometrically occurs both for a trivial rotation ($\theta=0$) and for a non trivial rotation whose unit vector is orthogonal to the reflection unit vector ($\theta \neq 0 $, $\hat{\bf r} \cdot \hat{\bf s} = 0$). Since a trivial rotation is the identity transformation its rotation axis is arbitrary and therefore even in the first subcase we set $\hat{\bf r} \cdot \hat{\bf s} = 0$. In both subcases, it is straightforward proving that the eigenvectors of $\tilde{R}$ are the vectors of Eqs.(\ref{uuu}), i.e. $\hat{\bf w}^{(n)} = \hat{\bf u}^{(n)}$, and that, for $\hat{\bf r} \cdot \hat{\bf s} = 0$, they reduce to
\begin{eqnarray} \label{bas1}
\hat{\bf u}^{(1)} &=& \hat{\bf r}, \nonumber \\
\hat{\bf u}^{(2)} &=& \cos \left(\frac{\theta}{2}\right) \hat{\bf s} \times \hat{\bf r} + \sin \left(\frac{\theta}{2}\right) \hat{\bf s}, \nonumber \\
\hat{\bf u}^{(3)} &=&  - \sin \left(\frac{\theta}{2}\right) \hat{\bf s} \times \hat{\bf r} + \cos \left(\frac{\theta}{2}\right) \hat{\bf s}.
\end{eqnarray}
Since for $\psi=0$ we have $\lambda_1=1$, $\lambda_2=1$, $\lambda_3 =-1$, from Eqs.(\ref{spectr}) we deduce that $V_3 = 0$ and $W_{mqn} = 0$ if only one or three of the indices $mqn$ are equal to $3$. Using Eqs.(\ref{vers1}), (\ref{vers2}) and (\ref{bas1}), from Eqs.(\ref{vec-tens}) we get after some algebra
\begin{eqnarray}
\epsilon_{imj} \overline{\varepsilon f_m} &=& V_1 \left(u_i^{(3)} u_j^{(2)} - u_i^{(2)} u_j^{(3)} \right) \nonumber \\
                                          &+& V_2 \left(u_i^{(1)} u_j^{(3)} - u_i^{(3)} u_j^{(1)} \right), \nonumber \\
\epsilon_{imn} \overline{\varepsilon f_m \partial_q f_n} &=& (W_{233}-W_{322}) u_i^{(1)} u_q^{(3)} \nonumber \\
                                                         &+& (W_{112}-W_{211}) u_i^{(3)} u_q^{(1)} \nonumber \\
                                                         &+& (W_{331}-W_{133}) u_i^{(2)} u_q^{(3)} \nonumber \\
                                                         &+& (W_{122}-W_{221}) u_i^{(3)} u_q^{(2)}.
\end{eqnarray}
Inserting these quantities into the definition of the chirality tensor of Eq.(\ref{kappa1}) and noting that $\epsilon_{qmn} \overline{\varepsilon f_m \partial_q f_n} = 0$ since $\hat{\bf u}^{(1)} \cdot \hat{\bf u}^{(3)} = \hat{\bf u}^{(2)} \cdot \hat{\bf u}^{(3)} = 0$ we get
\begin{eqnarray}
\kappa_{ij}= \eta \left( \kappa^{(13)} u_i^{(1)} u_j^{(3)} + \kappa^{(31)} u_i^{(3)} u_j^{(1)} \right. \nonumber \\
+ \left. \kappa^{(23)} u_i^{(2)} u_j^{(3)} + \kappa^{(32)} u_i^{(3)} u_j^{(2)} \right)
\end{eqnarray}
where $\kappa^{(13)} = k_0 \left(V_2 + W_{233} - W_{332}\right)$, $\kappa^{(31)} = k_0 \left(-V_2 + W_{112} - W_{211}\right)$
, $\kappa^{(23)} = k_0 \left(-V_1 + W_{331} - W_{133}\right)$ and $\kappa^{(32)} = k_0 \left(V_1 + W_{122} - W_{221}\right)$ are four scalars which are mutually independent since they depends on ten independent scalars. In matrix form the above chirality tensor reads
\begin{eqnarray}
\kappa = \eta \left( \kappa^{(13)} \hat{\bf u}^{(1)} \hat{\bf u}^{(3)T} + \kappa^{(31)} \hat{\bf u}^{(3)} \hat{\bf u}^{(1)T} \right. \nonumber \\
+ \left. \kappa^{(23)} \hat{\bf u}^{(2)} \hat{\bf u}^{(3)T} + \kappa^{(32)} \hat{\bf u}^{(3)} \hat{\bf u}^{(2)T} \right).
\end{eqnarray}

{\bf Case} $\psi=\frac{\pi}{2}, \frac{3\pi}{2}$. From Eq.(\ref{psi}) we deduce that this case geometrically occurs if the reflection and rotation units vectors and the rotation angle are such that $\sin^2 \left(\frac{\theta}{2}\right) ( \hat{\bf r}  \cdot \hat{\bf s})^2 = \frac{1}{2}$. In this case it is straightforward proving that the eigenvectors are
\begin{eqnarray} \label{bas2}
\hat{\bf w}^{(1)} &=& \frac{1}{\sqrt{2}} \left( \hat{\bf u}^{(1)} + i \hat{\bf u}^{(2)} \right), \nonumber \\
\hat{\bf w}^{(2)} &=& \frac{1}{\sqrt{2}} \left( \hat{\bf u}^{(1)} - i \hat{\bf u}^{(2)} \right), \nonumber \\
\hat{\bf w}^{(3)} &=& \hat{\bf u}^{(3)}.
\end{eqnarray}
Since for $\psi=\frac{\pi}{2},\frac{3\pi}{2}$ we have $\lambda_1=\pm i$, $\lambda_2=\mp i$, $\lambda_3 =-1$, from Eqs.(\ref{spectr}) we deduce that $V_1=V_2=V_3 = 0$ and $W_{mqn} = 0$ if it does not occur that one of the indices $mqn$ is equal to $3$ and the other two are both equal to $1$ or $2$. Using Eqs.(\ref{vers1}), (\ref{vers2}) and (\ref{bas2}), from Eqs.(\ref{vec-tens}) we get after some algebra
\begin{eqnarray}
\epsilon_{imj} \overline{\varepsilon f_m} &=& 0, \nonumber \\
\epsilon_{imn} \overline{\varepsilon f_m \partial_q f_n} &=&
        \frac{1}{2} \left( W_{322} - W_{233} - W_{113} + W_{311}  \right) \nonumber \\
        && \left( u_i^{(1)} u_q^{(2)} +  u_i^{(2)} u_q^{(1)} \right) \nonumber \\
        &+& \frac{i}{2} \left( W_{322} - W_{233} + W_{113} - W_{311}  \right) \nonumber \\
        && \left( u_i^{(1)} u_q^{(1)} +  u_i^{(2)} u_q^{(2)} \right).
\end{eqnarray}
Inserting these quantities into the definition of the chirality tensor of Eq.(\ref{kappa1}) we get
\begin{eqnarray}
\kappa_{ij} &=& \eta \left[ \kappa^{(s)} \left( u_i^{(1)} u_j^{(2)} + u_i^{(2)} u_j^{(1)} \right) \right. \nonumber \\
               && \left. + \kappa^{(a)} \left( u_i^{(1)} u_j^{(1)} - u_i^{(2)} u_j^{(2)} \right) \right]
\end{eqnarray}
where $\kappa^{(s)} = \frac{k_0}{2} \left( W_{322} - W_{233} - W_{113} + W_{311}  \right)$ and $\kappa^{(a)} = \frac{ik_0}{2} \left( W_{322} - W_{233} + W_{113} - W_{311}  \right)$ are two scalars which are mutually independent since they depends on four independent scalars. In matrix form the above chirality tensor reads
\begin{eqnarray}
\kappa &=& \eta \left[ \kappa^{(s)} \left( \hat{\bf u}^{(1)} \hat{\bf u}^{(2)T} + \hat{\bf u}^{(2)} \hat{\bf u}^{(1)T} \right) \right. \nonumber \\
       && \left. + \kappa^{(a)} \left( \hat{\bf u}^{(1)} \hat{\bf u}^{(1)T} - \hat{\bf u}^{(2)} \hat{\bf u}^{(2)T} \right) \right].
\end{eqnarray}

{\bf Case} $\psi=\frac{2\pi}{3}, \frac{4\pi}{3}$. From Eq.(\ref{psi}) we deduce that this case geometrically occurs if the reflection and rotation units vectors and the rotation angle are such that $\sin^2 \left(\frac{\theta}{2}\right) ( \hat{\bf r}  \cdot \hat{\bf s})^2 = \frac{3}{4}$. Since for $\psi=\frac{2\pi}{3}, \frac{4\pi}{3}$ we have $\lambda_1=\frac{1}{2}\left(-1\pm i \sqrt{3}\right)$, $\lambda_2=\frac{1}{2}\left(-1 \mp i \sqrt{3}\right)$, $\lambda_3 =-1$, from Eqs.(\ref{spectr}) we deduce that $V_1=V_2=V_3 = 0$ and $W_{mqn} = 0$ if it does not occur that $m=q=n=1$ or $m=q=n=2$ (i.e. only the components $W_{111}$ and $W_{222}$ survive). From Eqs.(\ref{vec-tens}) we directly get
\begin{eqnarray}
\epsilon_{imj} \overline{\varepsilon f_m} &=& 0, \nonumber \\
\epsilon_{imn} \overline{\varepsilon f_m \partial_q f_n}  &=& 0
\end{eqnarray}
so that from the definition of the chirality tensor of Eq.(\ref{kappa1}) we obtain
\begin{equation}
\kappa_{ij} = 0,
\end{equation}
i.e. the chirality tensor vanishes.

\section{Effective medium response of one-dimensional dielectric media}
Let us consider a metamaterial whose dielectric profile is $\varepsilon(Z)$, i.e. it is dependent on a single cartesian coordinate. The dielectric profile $\varepsilon(Z)$ and its inverse $\left[\varepsilon(Z)\right]^{-1}$ can be represented through the Fourier series
\begin{eqnarray} \label{1D-die}
\varepsilon(Z) &=& \sum_{n=-\infty}^{+\infty} e^{i n k_0 Z} a_n, \nonumber \\
\left[\varepsilon(Z)\right]^{-1} &=& \sum_{n=-\infty}^{+\infty} e^{i n k_0 Z} b_n
\end{eqnarray}
where $k_0 = 2\pi / \lambda$ and
\begin{eqnarray} \label{1D-fourier}
a_n &=& \frac{1}{\lambda} \int_{0}^{\lambda} dZ  e^{-i n k_0 Z} \varepsilon(Z) , \nonumber \\
b_n &=& \frac{1}{\lambda} \int_{0}^{\lambda} dZ  e^{-i n k_0 Z} \left[\varepsilon(Z)\right]^{-1}
\end{eqnarray}
are their Fourier coefficients. Due to the generality of the Fourier series representation, the situation we are here considering allows us to investigate even discontinuous dielectric distributions, a relevant situation occurring when the metamaterial is a stratified layered structure, without explicitly resorting to matching condition on the surfaces of the inclusions. Note also that
\begin{eqnarray} \label{averages}
a_0 &=& \overline{\varepsilon}, \nonumber \\
b_0 &=& \overline{\varepsilon^{-1}}.
\end{eqnarray}

For the one-dimensional dielectric distribution $\varepsilon(Z)$, the functions $f_j$ satisfying the first of Eqs.(\ref{eff-eqs}) are evidently dependent solely on $Z$, i.e. $f_j = f_j(Z)$, so that such equations yields
\begin{eqnarray} \label{syst1}
\frac{d}{dZ} \left( \varepsilon \frac{d f_j}{dZ} \right) &=& 0, \hspace{1cm} j = 1,2, \nonumber \\
\frac{d}{dZ} \left( \varepsilon \frac{d f_3}{dZ} \right) &=& - \frac{d \varepsilon}{d Z},
\end{eqnarray}
which have to be solved with the requirements that the functions $f_j$ have the same periodicity as $\varepsilon(Z)$ and have vanishing average $\overline{f_j}=0$. After an integration Eqs.(\ref{syst1}) become
\begin{eqnarray} \label{syst2}
\frac{d f_j}{dZ}  &=& \frac{C_j}{\varepsilon}, \hspace{1cm} j = 1,2, \nonumber \\
\frac{d f_3}{dZ}  &=& -1 + \frac{C_3}{\varepsilon}
\end{eqnarray}
where $C_1,C_2,C_3$ are constants. Averaging Eqs.(\ref{syst2}) over the metamaterial period, using Eqs.(\ref{averages}) and noting that the average of their left hand sides are zero since they are derivatives of periodic functions we get
\begin{eqnarray} \label{ave-syst2}
0  &=& C_j b_0 , \hspace{1cm} j = 1,2, \nonumber \\
0  &=& -1 + C_3 b_0.
\end{eqnarray}
which, assuming $b_0 \neq 0$, imply that $C_1=C_2=0$ and $C_3= 1/b_0$. Accordingly, from the first of Eqs.(\ref{syst2}) we obtain that $f_1$ and $f_2$ are constants so that, requiring their average to be zero, we get
\begin{equation} \label{1D-f12}
f_1 = f_2 = 0.
\end{equation}
Using the Fourier espansion of $\varepsilon^{-1}$ of the second of Eqs.(\ref{1D-die}), the second of Eqs.(\ref{syst2}) becomes
\begin{equation}
\frac{d f_3}{dZ} = \sum_{n=-\infty, n \neq 0}^{+\infty} e^{i n k_0 Z}  \frac{b_n}{b_0}
\end{equation}
which, after an integration and the requirement for $f_3$ to have vanishing average, yields
\begin{equation} \label{1D-f3}
f_3 =  \sum_{n=-\infty, n \neq 0}^{+\infty} e^{i n k_0 Z} \frac{b_n}{i n k_0 b_0}.
\end{equation}
Combining Eqs.(\ref{1D-f12}) and (\ref{1D-f3}) we obtain for the functions $f_j$
\begin{equation} \label{1D-fj}
f_j = \delta_{j3} \sum_{n=-\infty, n \neq 0}^{+\infty} e^{i n k_0 Z} \frac{b_n}{i n k_0 b_0}
\end{equation}
which are fully determined by the dielectric profile $\varepsilon(Z)$ and they are unique as a consequence of the chosen constraints (periodicity and vanishing average).

Using Eq.(\ref{1D-fj}), the first of Eqs.(\ref{eff-QP}) becomes
\begin{equation} \label{1D-Q}
Q_{ij} = \left(\sum_{m=-\infty}^{+\infty} e^{i m k_0 Z} a_m \right) \left( \delta_{ij} + \delta_{i3} \delta_{j3} \sum_{n=-\infty, n \neq 0}^{+\infty} e^{i n k_0 Z} \frac{b_n}{b_0} \right)
\end{equation}
which, inserted into the first of Eqs.(\ref{eff-tens}) and neglecting the $\eta^2$ contribution, yields
\begin{equation} \label{1D-eps1}
\varepsilon^{(eff)}_{ij} =  \delta_{ij} a_0 + \delta_{i3} \delta_{j3} \frac{1}{b_0} \sum_{n=-\infty, n \neq 0}^{+\infty} a_{-n} b_n.
\end{equation}
Averaging the relation $\varepsilon \varepsilon^{-1} = 1$ and using Eqs.(\ref{1D-die}) the relation $\sum_{n=-\infty}^{+\infty} a_{-n} b_n = 1$ easily follows so that $\sum_{n=-\infty, n \neq 0}^{+\infty} a_{-n} b_n = 1-a_0 b_0$ and Eq.(\ref{1D-eps1}) becomes
\begin{equation} \label{1D-eps2}
\varepsilon^{(eff)}_{ij} = (\delta_{i1} \delta_{j1}+\delta_{i2} \delta_{j2} ) a_0 + \delta_{i3} \delta_{j3} \frac{1}{b_0} .
\end{equation}
or in matrix form
\begin{equation} \label{1D-eps3}
\varepsilon^{(eff)} = \left(\begin{array}{ccc} \overline{\varepsilon} & 0 & 0 \\ 0  &  \overline{\varepsilon}  & 0 \\ 0 & 0 & \left[\overline{\varepsilon^{-1}}\right]^{-1} \end{array} \right)
\end{equation}
where Eqs.(\ref{averages}) have been used.

In order to evaluate the chirality tensor we note that Eqs.(\ref{1D-fj}) and the first of Eqs.(\ref{1D-die}) yield
\begin{eqnarray}
\overline{\varepsilon f_m} &=& \delta_{m3} \sum_{s=-\infty, s \neq 0}^{+\infty} \frac{a_{-s} b_s}{i k_0 b_0 s}, \nonumber \\
\overline{\varepsilon f_m \partial_q f_n} &=& \delta_{m3} \delta_{q3} \delta_{n3} \overline{\varepsilon f_3 \frac{d f_3}{dZ}}
\end{eqnarray}
so that, from Eq.(\ref{kappa1}) we obtain
\begin{equation} \label{1D-kappa}
\kappa_{ij} =  \epsilon_{ij3} \eta \kappa_0
\end{equation}
where
\begin{equation} \label{1D-kappa0}
\kappa_0 = \frac{i }{b_0} \sum_{s=-\infty, s \neq 0}^{+\infty} \frac{a_{-s} b_s}{s}.
\end{equation}
In matrix form, Eq.(\ref{1D-kappa}) reads
\begin{equation} \label{1D-eps3}
\kappa =   \left(\begin{array}{ccc} 0 & \eta \kappa_0 & 0 \\ -\eta \kappa_0  &  0  & 0 \\ 0 & 0 & 0 \end{array} \right).
\end{equation}
Inserting Eqs.(\ref{1D-fourier}) into Eq.(\ref{1D-kappa0}) we obtain
\begin{equation} \label{1D-kappa0-1}
\kappa_0 = \frac{i }{b_0 \lambda^2} \int_0^\lambda dZ_1 \int_0^\lambda dZ_2 \frac{\varepsilon(Z_1)}{\varepsilon(Z_2)} \sum_{s=-\infty, s \neq 0}^{+\infty} \frac{e^{i 2\pi s \left(\frac{Z1-Z2}{\lambda}\right)}}{s}.
\end{equation}
After noting that the even contribution in the numerator within the series does not contribute and using the relation $\sum_{s=1}^{+\infty} \frac{sin(2 \pi s \xi)}{s} = \pi \left[-\xi + \frac{1}{2} \textrm{sign}(\xi) \right]$ valid for $| \xi | < 1$, Eq.(\ref{1D-kappa0-1}) yields
\begin{eqnarray} \label{1D-kappa0-1}
\kappa_0 &=& \left[\overline{\varepsilon^{-1}}\right]^{-1} \frac{2 \pi }{ \lambda^2} \int_0^\lambda dZ_1 \int_0^\lambda dZ_2 \frac{\varepsilon(Z_1)}{\varepsilon(Z_2)} \left[ \left(\frac{Z_1-Z_2}{\lambda}\right)  \right. \nonumber \\ &&-
\left. \frac{1}{2} \textrm{sign} \left(\frac{Z_1-Z_2}{\lambda}\right) \right]
\end{eqnarray}
where the first of Eqs.(\ref{averages}) has been used.



\begin{thebibliography} {aa}
\bibitem{OBrien} S. O{'}Brien, and J. B. Pendry, J. Phys. Condens. Matter \textbf{14}, 6383 (2002).
\bibitem{Smith1} D. R. Smith, S. Schultz, P. Markos and C. M. Soukoulis, Phys. Rev. B \textbf{65}, 195104 (2002).
\bibitem{Smith2} D. R. Smith, D. C. Vier, T. Koschny, and C. M. Soukoulis, Phys. Rev. E \textbf{71}, 036617 (2005).
\bibitem{Chennn} X. Chen, B. Wu, J. Kong, and T. M. Grzegorczyk, Phys. Rev. E \textbf{71}, 046610 (2005).
\bibitem{Menzel} C. Menzel, C. Rockstuhl, T. Paul, and F. Lederer, Phys. Rev. B \textbf{77} 195328 (2008).
\bibitem{Plumm1} E. Plum, J. Zhou, J. Dong, V. A. Fedotov, T. Koschny, C. M. Soukoulis, and N. I. Zheludev  Phys. Rev. B
                 \textbf{79} 035407 (2009).
\bibitem{Simovs} C. R. Simovski, and S. A. Tretyakov, Phys. Rev. B \textbf{75}, 195111 (2007).
\bibitem{Karama} T. D. Karamanos, S. D. Assimonis, A. I. Dimitriadis, and N. V. Kantartzis, Photon. Nanostruct. Fundam. Appl.
                 \textbf{12}, 291 (2014).
\bibitem{Pendry} J. B. Pendry, A. J. Holden, D. J. Robbins, and W. J. Stewart IEEE Trans. Microw. Theory Tech. \textbf{47}, 2075 (1999).
\bibitem{Smith3} D. R. Smith, D. C. Vier, N. Kroll, and S. Schultz, Appl. Phys. Lett. \textbf{77}, 2246 (2000).
\bibitem{Smith4} D. R. Smith, and J. B. Pendry, J. Opt. Soc. Am. B \textbf{23}, 391 (2006).
\bibitem{Tsuker} I. Tsukerman, J. Opt. Soc. Am. B \textbf{28}, 577 (2011).
\bibitem{Ouchet} O. Ouchetto, H. Ouchetto, S. Zouhdi, and A. Sekkaki, IEEE Trans. Antennas Propag. \textbf{61}, 4214 (2013).
\bibitem{Ishima} A. Ishimaru, S. Lee, Y. Kuga, and V. Jandhyala, IEEE Trans. Antenna Propag. \textbf{51}, 2551 (2003).
\bibitem{Belov1} P. Belov, and C R. Simovski, Phys. Rev. E \textbf{72}, 026615 (2005).
\bibitem{Lambbb} W. Lamb, D. M. Wood, and N. W. Ashcroft, Phys. Rev. B \textbf{21}, 2248 (1980).
\bibitem{Aluuuu} A. Alu, Phys. Rev. B \textbf{84}, 075153 (2011).
\bibitem{Silvei} M. G. Silveirinha, Phys. Rev. B \textbf{75}, 115104 (2007).
\bibitem{Taoooo} R. Tao, Z. Chen, and P. Sheng, Phys. Rev. B \textbf{41}, 2417 (1990).
\bibitem{Dattaa} S. Datta, C. T. Chan, K. M. Ho, and C. M. Soukoulis, Phys. Rev. B \textbf{48}, 14936 (1993).
\bibitem{Halevi} P. Halevi, A. A. Krokhin, and J. Arriaga, Phys. Rev. Lett. \textbf{82}, 719 (1999).
\bibitem{Ortizz} G. P. Ortiz, B. E. Martínez-Zérega, B. S. Mendoza, and W. L. Mochán, Phys. Rev. B \textbf{79}, 245132 (2009).
\bibitem{Reyes1} J. A. Reyes-Avendaño, U. Algredo-Badillo, P. Halevi, and F. Pérez-Rodríguez, New J. Phys. \textbf{13}, 073041 (2011).
\bibitem{PerezH} J. S. Perez-Huerta, G. P. Ortiz, B. S. Mendoza and W. L. Mochán, New J. Phys. \textbf{15},  043037 (2013)
\bibitem{Reyes2} J. A. Reyes-Avendaño, M. P. Sampedro, E. Juárez-Ruiz, and F Pérez-Rodríguez, J. Opt. \textbf{16}, 065102 (2014).
\bibitem{Felba0} D. Felbacq, and G. Bouchitté, Phys. Rev. Lett. \textbf{94}, 183902 (2005).
\bibitem{Ouche1} O. Ouchetto, S. Zouhdi, A. Bossavit, G. Griso, B. Miara, and A. Razek, J. Mat. Proc. Techn. \textbf{181}, 225 (2007).
\bibitem{Felbac} D. Felbacq, G. Bouchitté, B. Guizal, and A. Moreau, J. Nanophoton. \textbf{2}, 023501 (2008).
\bibitem{Chenn2} Y. Chen, and R. Lipton, Photon. Nanostruct. Fundam. Appl. \textbf{11}, 442 (2013).
\bibitem{Rizza1} C. Rizza, and A. Ciattoni, Phys. Rev. Lett. \textbf{110}, 143901 (2013).
\bibitem{Rizza2} C. Rizza, and A. Ciattoni, Opt. Lett. \textbf{38}, 3658 (2013).
\bibitem{Rizza3} C. Rizza, E. Palange, and A. Ciattoni, Photon. Res. \textbf{2}, 121 (2014).
\bibitem{Belov2} P. A. Belov, R. Marque´s, S. I. Maslovski, I. S. Nefedov, M. Silveirinha, C. R. Simovski, and S. A. Tretyakov, Phys. Rev. B \textbf{67}, 113103 (2003).
\bibitem{Gerken} M. Gerken, and D. A. B. Miller, Appl. Opt. \textbf{42}, 1330 (2003).
\bibitem{Silve2} M. G. Silveirinha, Phys. Rev. E \textbf{73}, 046612 (2006).
\bibitem{Elserr} J. Elser, V. A. Podolskiya, I. Salakhutdinov, and I. Avrutsky, Appl. Phys. Lett. \textbf{90}, 191109 (2007).
\bibitem{Pollar} R. J. Pollard, A. Murphy, W. R. Hendren, P. R. Evans, R. Atkinson, G. A. Wurtz, A.V. Zayats, and V. A. Podolskiy, Phys. Rev. Lett. \textbf{102}, 127405 (2009).
\bibitem{Chebyk} A. V. Chebykin, A. A. Orlov, A. V. Vozianova, S. I. Maslovski, Yu. S. Kivshar, and P. A. Belov, Phys. Rev. B \textbf{84}, 115438 (2011).
\bibitem{Orlovv} A. A. Orlov, P. M. Voroshilov, P. A. Belov, and Y. S. Kivshar, Phys. Rev. B \textbf{84},  045424 (2011).

\bibitem{Serdyu} A.N. Serdyukov, I.V. Semchenko, S.A. Tretyakov, A. Sihvola, \emph{Electromagnetics of Bi-Anisotropic Materials: Theory and Applications.} (Gordon and Breach Science Publishers, Amsterdam 2001).
\bibitem{Menze1} C. Menzel, T. Paul, C. Rockstuhl, T. Pertsch, S. Tretyakov, and Falk Lederer, Phys. Rev. B \textbf{81}, 035320 (2010)
\bibitem{Iwanag} M. Iwanaga, Phys. Stat. Sol. (b) \textbf{245}, 2684 (2008).
\bibitem{Tserke} C. Tserkezis, N. Papanikolaou, G. Gantzounis, and N. Stefanou, Phys. Rev. B \textbf{78}, 165114 (2008).
\bibitem{Chernn} R. L. Chern, and D. Felbacq, Phys. Rev. B \textbf{79}, 075118 (2009).
\bibitem{Ginnnn} J. C. Ginn, and I. Brener, Phys. Rev. Lett. \textbf{108}, 097402 (2012).
\bibitem{Jeli01} L. Jelinek, R. Marqu\'{e}s, F. Mesa, J. D. Baena, Phys. Rev. B \textbf{77}, 205110 (2008).
\bibitem{Gansel} J. K. Gansel, M. Thiel, M. S. Rill, M. Decker, K. Bade, V. Saile, G. von Freymann,
S. Linden, M. Wegener, Science \textbf{325}, 1513 (2009).
\bibitem{Plum01}  E. Plum, J. Zhou, J. Dong, V. A. Fedotov, T. Koschny, C. M. Soukoulis, N. I. Zheludev, Phys. Rev. B \textbf{79}, 035407 (2009).
\bibitem{Wang01} B. Wang, J. Zhou, T. Koschny, C. M. Soukoulis, Appl. Phys. Lett. \textbf{94}, 151112 (2009)
\bibitem{Papako} A. Papakostas, A. Potts, D.M. Bagnall, S. L. Prosvirnin, H. J. Coles, and N. I. Zheludev Phys. Rev. Lett. \textbf{90}, 107404-1 (2003).
\bibitem{Fedoto} V. A. Fedotov, P. L. Mladyonov, S. L. Prosvirnin, A. V. Rogacheva, Y. Chen, N. I. Zheludev, Phys. Rev. Lett. \textbf{97}, 167401 (2006).
\bibitem{Singhh} R. Singh, E. Plum, C. Menzel, C. Rockstuhl, A. K. Azad, R. A. Cheville, F. Lederer, W. Zhang, N. I. Zheludev, Phys. Rev. B \textbf{80}, 153104 (2009).
\bibitem{Zhangg} S. Zhang, Y. S. Park, J. Li, X. Lu, W. Zhang, X. Zhang, Phys. Rev. Lett. \textbf{102}, 023901 (2009).
\bibitem{Yeeeee} Y. Ye, S. He, Appl. Phys. Lett. \textbf{96}, 203501 (2010).
\bibitem{Menze2} C. Menzel, C. Helgert, C. Rockstuhl, E. B. Kley, A. Tunnermann, T. Pertsch, and F. Lederer, Phys. Rev. Lett. \textbf{104}, 253902 (2010).
\bibitem{Liiiii} Z. Li, M. Mutlu, E. Ozbay, J. Opt. \textbf{15}, 023001 (2013).
\bibitem{Zhaooo} R. Zhao, J. Zhou, Th. Koschny, E. N. Economou, C. M. Soukoulis, Phys. Rev. Lett. \textbf{103}, 103602 (2009).

\bibitem{Trety3} S. Tretyakov, A. Sihvola, and L. Jylha, Photon. Nanostruct. Fundam. Appl. \textbf{3}, 107 (2005).
\bibitem{Chern1} R. L. Chern, J. Phys. D: Appl. Phys. \textbf{46}, 125307 (2013).
\bibitem{Pendr2} J. B. Pendry, Science \textbf{306}, 1353-1355 (2004).
\bibitem{Trety1} S. Tretyakov, I. Nefedov, A. Sihvola, S. Maslovski, C. Simovski, J. Electromagn. Waves Appl. \textbf{17}, 695 (2003).
\bibitem{Trety2} S. Tretyakov, A. Sihvola, L. Jylha, Photon. Nanostruct. Fundam. Appl. \textbf{3}, 107 (2005).
\bibitem{Monzon} C. Monzon, D. Forester, Phys. Rev. Lett. \textbf{95}, 123904 (2005).
\bibitem{Sander} J. A. Sanders and F. Verhulst, Averaging Methods in Nonlinear Dynamical Systems (Springer-Verlag, Berlin,
1985).
\bibitem{Landau} L.D. Landau and E.M. Lifshitz, \emph{Electrodynamics of Continuous Media.} (Pergamon Press, 1960).
\bibitem{Capoli} F. Capolino, \emph{Theory and Phenomena of Metamaterials} (Taylor and Francis Group, Boca Raton, FL, 2009).

\end{thebibliography}
\end{document}